\documentclass[twocolumn,twocolappendix,times]{aastex701}
\usepackage{amsmath,amssymb}
\usepackage{CJKutf8}
\usepackage{enumitem}
\usepackage{float}
\usepackage{lineno}
\usepackage{lipsum}
\usepackage{subfig}
\usepackage{times}
\usepackage{wasysym}
\usepackage[normalem]{ulem}
\linenumbers

\usepackage{silence}
\newcommand{\fMg}{f_\mathrm{Mg \, \textsc{ii}}}
\newcommand{\NMg}{N^\mathrm{ion}_\mathrm{Mg \, \textsc{ii}}}
\newcommand{\nMg}{n^\mathrm{ion}_\mathrm{Mg \, \textsc{ii}}}
\newcommand{\WMg}{W_\mathrm{Mg \, \textsc{ii}}}

\newcommand{\clin}{c_\mathrm{line}}
\newcommand{\Fcon}{F_\mathrm{cont}}
\newcommand{\Fqso}{F_\mathrm{quasar}}
\newcommand{\fcgm}{f_\mathrm{cool}}
\newcommand{\fdrt}{f_\mathrm{DR}}
\newcommand{\flin}{f_\mathrm{line}}
\newcommand{\fqso}{f_\mathrm{quasar}}
\newcommand{\ftrn}{f_\mathrm{trans}}
\newcommand{\lamr}{\lambda_\mathrm{rf}}
\newcommand{\Mvir}{M_\mathrm{vir}}
\newcommand{\rhcl}{\rho_\mathrm{cool}}
\newcommand{\rvir}{r_\mathrm{vir}}
\newcommand{\vlos}{v_\mathrm{los}}
\newcommand{\wpgc}{w^\mathrm{gc}_p}
\newcommand{\wpga}{w^\mathrm{ga}_p}

\newcommand{\xrgc}{\xi^r_\mathrm{gc}}
\newcommand{\xrOh}{\xi^r_\mathrm{1h}}
\newcommand{\xrTh}{\xi^r_\mathrm{2h}}
\newcommand{\xsga}{\xi^s_\mathrm{ga}}
\newcommand{\xsgc}{\xi^s_\mathrm{gc}}
\newcommand{\xsOh}{\xi^s_\mathrm{1h}}
\newcommand{\xsTh}{\xi^s_\mathrm{2h}}
\newcommand{\zeff}{z_\mathrm{eff}}

\begin{document}

\title{A Knuckle in the Finger of God: Possible Signature of Cool Circumgalactic Outflows around DESI LRGs}
\shorttitle{Kinematics of Cool CGM}

\author[orcid=0009-0009-6638-8830, gname=Yunzhe, sname='Gu']{Yunzhe Gu \begin{CJK*}{UTF8}{gbsn}\normalfont(顾韫哲)\end{CJK*}}
\affiliation{State Key Laboratory of Dark Matter Physics \& Tsung-Dao Lee Institute, Shanghai Jiao Tong University, Shanghai 200240, China}
\email{y-z.gu@sjtu.edu.cn}

\correspondingauthor{Ying Zu}
\author[orcid=0000-0001-6966-6925, gname=Ying, sname='Zu']{Ying Zu \begin{CJK*}{UTF8}{gbsn}\normalfont(祖颖)\end{CJK*}}
\affiliation{State Key Laboratory of Dark Matter Physics \& Tsung-Dao Lee Institute, Shanghai Jiao Tong University, Shanghai 200240, China}
\affiliation{Department of Astronomy, School of Physics and Astronomy, Shanghai Jiao Tong University, Shanghai 200240, China}
\email[show]{yingzu@sjtu.edu.cn}

\author[orcid=0000-0003-0822-452X, gname='Jessica Nicole', sname='Aguilar']{Jessica Nicole Aguilar}
\affiliation{Lawrence Berkeley National Laboratory, 1 Cyclotron Road, Berkeley, CA 94720, USA}
\email{jaguilar@lbl.gov}

\author[orcid=0000-0001-6098-7247, gname=Steven, sname='Ahlen']{Steven Ahlen}
\affiliation{Department of Physics, Boston University, 590 Commonwealth Avenue, Boston, MA 02215 USA}
\email{ahlen@bu.edu}

\author[orcid=0000-0001-7600-5148, gname=Eric, sname='Armengaud']{Eric Armengaud}
\affiliation{IRFU, CEA, Universit\'{e} Paris-Saclay, F-91191 Gif-sur-Yvette, France}
\email{eric.armengaud@cea.fr}

\author[orcid=0000-0001-5998-3986, gname=Alejandro, sname='Aviles']{Alejandro Aviles}
\affiliation{Instituto Avanzado de Cosmolog\'{\i}a A.~C., San Marcos 11 - Atenas 202. Magdalena Contreras. Ciudad de M\'{e}xico C.~P.~10720, M\'{e}xico}
\affiliation{Instituto de Ciencias F\'{\i}sicas, Universidad Nacional Aut\'onoma de M\'exico, Av. Universidad s/n, Cuernavaca, Morelos, C.~P.~62210, M\'exico}
\email{aviles@icf.unam.mx}

\author[orcid=0000-0003-0467-5438, gname=Florian, sname='Beutler']{Florian Beutler}
\affiliation{Institute for Astronomy, University of Edinburgh, Royal Observatory, Blackford Hill, Edinburgh EH9 3HJ, UK}
\email{florian.beutler@ed.ac.uk}

\author[orcid=0000-0001-9712-0006, gname=Davide, sname='Bianchi']{Davide Bianchi}
\affiliation{Dipartimento di Fisica ``Aldo Pontremoli'', Universit\`a degli Studi di Milano, Via Celoria 16, I-20133 Milano, Italy}
\affiliation{INAF-Osservatorio Astronomico di Brera, Via Brera 28, 20122 Milano, Italy}
\email{davide.bianchi1@unimi.it}

\author[orcid=0000-0002-8458-5047, gname=David, sname='Brooks']{David Brooks}
\affiliation{Department of Physics \& Astronomy, University College London, Gower Street, London, WC1E 6BT, UK}
\email{david.brooks@ucl.ac.uk}

\author[orcid=0000-0003-3044-5150, gname=Aurelio, sname='Carnero Rosell']{Aurelio Carnero Rosell}
\affiliation{Departamento de Astrof\'{\i}sica, Universidad de La Laguna (ULL), E-38206, La Laguna, Tenerife, Spain}
\affiliation{Instituto de Astrof\'{\i}sica de Canarias, C/ V\'{\i}a L\'{a}ctea, s/n, E-38205 La Laguna, Tenerife, Spain}
\email{acarnero@iac.es}

\author[orcid=0000-0001-8996-4874, gname=Edmond, sname='Chaussidon']{Edmond Chaussidon}
\affiliation{Lawrence Berkeley National Laboratory, 1 Cyclotron Road, Berkeley, CA 94720, USA}
\email{echaussidon@lbl.gov}

\author[orcid=0000-0002-5024-6987, gname=Todd, sname='Claybaugh']{Todd Claybaugh}
\affiliation{Lawrence Berkeley National Laboratory, 1 Cyclotron Road, Berkeley, CA 94720, USA}
\email{tmclaybaugh@lbl.gov}

\author[orcid=0000-0002-2169-0595, gname=Andrei, sname='Cuceu']{Andrei Cuceu}
\affiliation{Lawrence Berkeley National Laboratory, 1 Cyclotron Road, Berkeley, CA 94720, USA}
\email{acuceu@lbl.gov}

\author[orcid=0000-0002-1769-1640, gname=Axel, sname='de la Macorra']{Axel de la Macorra}
\affiliation{Instituto de F\'{\i}sica, Universidad Nacional Aut\'{o}noma de M\'{e}xico, Cd. de M\'{e}xico C.~P.~04510, M\'{e}xico}
\email{macorra@fisica.unam.mx}

\author[orcid=0000-0002-6397-4457, gname=Peter, sname='Doel']{Peter Doel}
\affiliation{Department of Physics \& Astronomy, University College London, Gower Street, London, WC1E 6BT, UK}
\email{apd@star.ucl.ac.uk}

\author[orcid=0000-0002-2890-3725, gname='Jaime E.', sname='Forero-Romero']{Jaime E. Forero-Romero}
\affiliation{Departamento de F\'isica, Universidad de los Andes, Cra. 1 No. 18A-10, Edificio Ip, CP 111711, Bogot\'a, Colombia}
\affiliation{Observatorio Astron\'omico, Universidad de los Andes, Cra. 1 No. 18A-10, Edificio H, CP 111711 Bogot\'a, Colombia}
\email{je.forero@uniandes.edu.co}

\author[orcid=0000-0001-9632-0815, gname=Enrique, sname='Gaztanaga']{Enrique Gazta\~naga}
\affiliation{Institute of Cosmology \& Gravitation, University of Portsmouth, Portsmouth PO1 3FX, UK}
\affiliation{Institut d'Estudis Espacials de Catalunya (IEEC), 08034 Barcelona, Spain}
\affiliation{Institute of Space Sciences (ICE, CSIC), 08193 Barcelona, Spain}
\email{gaztanaga@gmail.com}

\author[orcid=0000-0003-3142-233X, gname=Satya, sname='Gontcho A Gontcho']{Satya Gontcho A Gontcho}
\affiliation{Department of Astronomy, University of Virginia, Charlottesville, VA 22904, USA}
\email{satya@virginia.edu}

\author[orcid=0000-0003-0825-0517, gname=Gaston, sname='Gutierrez']{Gaston Gutierrez}
\affiliation{Fermi National Accelerator Laboratory, PO Box 500, Batavia, IL 60510, USA}
\email{gaston@fnal.gov}

\author[orcid=0000-0002-2312-3121, gname=Boryana, sname='Hadzhiyska']{Boryana Hadzhiyska}
\affiliation{Institute of Astronomy, University of Cambridge, Madingley Road, Cambridge CB3 0HA, UK}
\affiliation{Kavli Institute for Cosmology Cambridge, Madingley Road, Cambridge CB3 0HA, UK}
\email{boryanah@alumni.princeton.edu}

\author[orcid=0000-0002-9136-9609, gname='Hiram K.', sname='Herrera-Alcantar']{Hiram K. Herrera-Alcantar}
\affiliation{Institut d'Astrophysique de Paris, 98 bis boulevard Arago, 75014 Paris, France}
\affiliation{IRFU, CEA, Universit\'{e} Paris-Saclay, F-91191 Gif-sur-Yvette, France}
\email{herrera.301@osu.edu}

\author[orcid=0000-0002-6024-466X, gname=Mustapha, sname='Ishak']{Mustapha Ishak}
\affiliation{Department of Physics, The University of Texas at Dallas, 800 W. Campbell Rd., Richardson, TX 75080, USA}
\email{mishak@utdallas.edu}

\author[orcid=0000-0002-5652-8870, gname=Tanveer, sname='Karim']{Tanveer Karim}
\affiliation{Department of Astronomy \& Astrophysics, University of Toronto, Toronto, ON M5S 3H4, Canada}
\email{tanveer.karim@utoronto.ca}

\author[gname=Robert, sname='Kehoe']{Robert Kehoe}
\affiliation{Department of Physics, Southern Methodist University, 3215 Daniel Avenue, Dallas, TX 75275, USA}
\email{kehoe@physics.smu.edu}

\author[orcid=0000-0001-6356-7424, gname=Anthony, sname='Kremin']{Anthony Kremin}
\affiliation{Lawrence Berkeley National Laboratory, 1 Cyclotron Road, Berkeley, CA 94720, USA}
\email{akremin@lbl.gov}

\author[orcid=0000-0002-6731-9329, gname=Claire, sname='Lamman']{Claire Lamman}
\affiliation{Center for Cosmology and AstroParticle Physics (CCAPP), Ohio State University, Columbus, OH 43210, USA}
\affiliation{Department of Astronomy, The Ohio State University, Columbus, OH 43210, USA}
\affiliation{Department of Physics, The Ohio State University, Columbus, OH 43210, USA}
\email{lamman.1@osu.edu}

\author[orcid=0000-0003-1838-8528, gname=Martin, sname='Landriau']{Martin Landriau}
\affiliation{Lawrence Berkeley National Laboratory, 1 Cyclotron Road, Berkeley, CA 94720, USA}
\email{mlandriau@lbl.gov}

\author[orcid=0000-0001-7178-8868, gname=Laurent, sname='Le Guillou']{Laurent Le Guillou}
\affiliation{Sorbonne Universit\'{e}, CNRS/IN2P3, Laboratoire de Physique Nucl\'{e}aire et de Hautes Energies (LPNHE), FR-75005 Paris, France}
\email{llg@lpnhe.in2p3.fr}

\author[orcid=0000-0002-1125-7384, gname=Aaron, sname='Meisner']{Aaron Meisner}
\affiliation{NSF NOIRLab, 950 N. Cherry Ave., Tucson, AZ 85719, USA}
\email{aaron.meisner@noirlab.edu}

\author[orcid=0000-0002-6610-4836, gname=Ramon, sname='Miquel']{Ramon Miquel}
\affiliation{Instituci\'{o} Catalana de Recerca i Estudis Avan\c{c}ats, Passeig de Llu\'{\i}s Companys, 23, 08010 Barcelona, Spain}
\affiliation{Institut de F\'{i}sica d’Altes Energies (IFAE), The Barcelona Institute of Science and Technology, Edifici Cn, Campus UAB, 08193 Bellaterra (Barcelona), Spain}
\email{rmiquel@ifae.es}

\author[orcid=0000-0002-2733-4559, gname=John, sname='Moustakas']{John Moustakas}
\affiliation{Department of Physics and Astronomy, Siena University, 515 Loudon Road, Loudonville, NY 12211, USA}
\email{jmoustakas@siena.edu}

\author[orcid=0000-0001-9070-3102, gname=Seshadri, sname='Nadathur']{Seshadri Nadathur}
\affiliation{Institute of Cosmology and Gravitation, University of Portsmouth, Dennis Sciama Building, Portsmouth, PO1 3FX, UK}
\email{seshadri.nadathur@port.ac.uk}

\author[orcid=0000-0001-8684-2222, gname='Jeffrey A.', sname='Newman']{Jeffrey A. Newman}
\affiliation{University of Pittsburgh, 100 Allen Hall, 3941 O'Hara Street, Pittsburgh, PA 15260, USA}
\email{janewman@pitt.edu}

\author[orcid=0000-0002-4637-2868, gname=Enrique, sname='Paillas']{Enrique Paillas}
\affiliation{Instituto de Estudios Astrof\'isicos, Facultad de Ingenier\'ia y Ciencias, Universidad Diego Portales, Av. Ej\'ercito Libertador 441, Santiago, Chile}
\email{enrique.paillas@udp.cl}

\author[orcid=0000-0002-0644-5727, gname='Will J.', sname='Percival']{Will J. Percival}
\affiliation{Department of Physics and Astronomy, University of Waterloo, 200 University Ave W, Waterloo, ON N2L 3G1, Canada}
\affiliation{Perimeter Institute for Theoretical Physics, 31 Caroline St. North, Waterloo, ON N2L 2Y5, Canada}
\affiliation{Waterloo Centre for Astrophysics, University of Waterloo, 200 University Ave W, Waterloo, ON N2L 3G1, Canada}
\email{will.percival@uwaterloo.ca}

\author[orcid=0000-0001-7145-8674, gname=Francisco, sname='Prada']{Francisco Prada}
\affiliation{Instituto de Astrof\'{i}sica de Andaluc\'{i}a (CSIC), Glorieta de la Astronom\'{i}a, s/n, E-18008 Granada, Spain}
\affiliation{Instituto de Astrof\'{\i}sica de Canarias, C/ V\'{\i}a L\'{a}ctea, s/n, E-38205 La Laguna, Tenerife, Spain}
\email{fprada@iaa.es}

\author[orcid=0000-0001-6979-0125, gname=Ignasi, sname='Perez-Rafols']{Ignasi P\'{e}rez-R\`{a}fols}
\affiliation{Departament de F\'isica, EEBE, Universitat Polit\`ecnica de Catalunya, c/Eduard Maristany 10, E-08930 Barcelona, Spain}
\email{ignasi.perez.rafols@upc.edu}

\author[orcid=0000-0002-3500-6635, gname=Corentin, sname='Ravoux']{Corentin Ravoux}
\affiliation{Universit\'{e} Clermont-Auvergne, CNRS, LPCA, 63000 Clermont-Ferrand, France}
\email{corentin.ravoux@clermont.in2p3.fr}

\author[gname=Graziano, sname='Rossi']{Graziano Rossi}
\affiliation{Department of Physics and Astronomy, Sejong University, 209 Neungdong-ro, Gwangjin-gu, Seoul 05006, Republic of Korea}
\email{graziano@sejong.ac.kr}

\author[orcid=0000-0002-0394-0896, gname=Rossana, sname='Ruggeri']{Rossana Ruggeri}
\affiliation{Queensland University of Technology, School of Chemistry \& Physics, George St, Brisbane 4001, Australia}
\email{rossana.ruggeri@qut.edu.au}

\author[orcid=0000-0002-9646-8198, gname=Eusebio, sname='Sanchez']{Eusebio Sanchez}
\affiliation{CIEMAT, Avenida Complutense 40, E-28040 Madrid, Spain}
\email{eusebio.sanchez@ciemat.es}

\author[orcid=0000-0002-0408-5633, gname=Christoph, sname='Saulder']{Christoph Saulder}
\affiliation{Max Planck Institute for Extraterrestrial Physics, Gie\ss enbachstra\ss e 1, 85748 Garching, Germany}
\email{csaulder@mpe.mpg.de}

\author[orcid=0000-0002-5042-5088, gname=David, sname='Schlegel']{David Schlegel}
\affiliation{Lawrence Berkeley National Laboratory, 1 Cyclotron Road, Berkeley, CA 94720, USA}
\email{djschlegel@lbl.gov}

\author[orcid=0000-0002-3461-0320, gname='Joseph Harry', sname='Silber']{Joseph Harry Silber}
\affiliation{Lawrence Berkeley National Laboratory, 1 Cyclotron Road, Berkeley, CA 94720, USA}
\email{jhsilber@lbl.gov}

\author[orcid=0000-0003-1704-0781, gname=Gregory, sname='Tarle']{Gregory Tarl\'{e}}
\affiliation{University of Michigan, 500 S. State Street, Ann Arbor, MI 48109, USA}
\email{gtarle@umich.edu}

\author[gname='Benjamin Alan', sname='Weaver']{Benjamin Alan Weaver}
\affiliation{NSF NOIRLab, 950 N. Cherry Ave., Tucson, AZ 85719, USA}
\email{benjamin.weaver@noirlab.edu}

\shortauthors{Gu, Zu, et al.}

\begin{abstract}
    The kinematics of cool, metal-enriched gas in the circumgalactic medium (CGM) provide a critical diagnostic of its physical origins. We present a novel spectral stacking technique to measure the redshift-space distortion (RSD) of \ion{Mg}{2}-absorbing clouds around luminous red galaxies (LRGs), using over two million quasar sightlines from DESI Data Release 2. Rather than employing traditional pair-counting methods based on discrete absorbers, we directly transform stacked absorption-line profiles into a redshift-space galaxy-cloud cross-correlation function, $\xsgc$. This stacked RSD measurement incorporates weak absorption features that fall below the detection threshold of discrete absorber catalogs, revealing a distinct ``knuckle'' feature superimposed on the standard Finger-of-God effect. To interpret the observed DESI RSD signal, we develop a bimodal skewed-$t$ (BST) velocity distribution model that captures the complex 3D motion of cool clouds around mock LRGs in the TNG50 simulation. Applying this flexible BST model to the stacked $\xsgc$ measurement, we find that the knuckle corresponds to a weak but statistically significant outflowing component with a characteristic radial velocity of $\sim 459 \, \mathrm{km \, s^{-1}}$. This inferred outflow velocity declines monotonically with increasing post-starburst age, consistent with a ``tired wind'' scenario in which outflows launched after the last star-forming episode gradually decelerate within the halo potential. Our method offers a promising pathway to unlocking the complex kinematics of CGM gas through stacked RSD measurements from absorption spectroscopy.
\end{abstract}


% The "Package hyperref: Ignoring empty anchor." error is due to the unnumbered footnote of \email[show]. I modified the .cls file to fix it. Now the compilation should be error-free.
\section{Introduction}
\label{sec:intro}

The circumgalactic medium (CGM) of luminous red galaxies (LRGs) contains a surprisingly large amount of cool ($\sim 10^4 \, \mathrm{K}$), metal-enriched gas of unknown origin~\citep{Huang2016, Chen2018, Zahedy2019, Berg2019}. While previous studies of the CGM have primarily focused on the spatial distribution of cool clouds~\citep{Chen2010b, Bordoloi2011, Nielsen2013, Werk2013}, kinematics provide the crucial evidence needed to uncover the physical origin of these clouds (e.g., inflow vs.\ outflow). In this paper, we develop a novel spectroscopic stacking technique to measure the redshift-space distortion (RSD) of \ion{Mg}{2} gas in the LRG halos, using over two million quasar sightlines from the Dark Energy Spectroscopic Instrument~\citep[DESI;][]{Levi2013, DESI2016a, DESI2016b, DESI2022} Data Release 2~\citep[DR2;][]{Karim2025a, Karim2025b, DESI2026}. By interpreting the stacked RSD with an analytic model informed by the IllustrisTNG hydrodynamical simulation~\citep{Pillepich2018, Nelson2018}, we hope to infer the 3D velocity distribution of the \ion{Mg}{2} gas and determine its physical origin.

Traditionally, kinematic studies of \ion{Mg}{2} absorbers are primarily based on the individual line profiles, which can be related to the line-of-sight (LOS) velocity distribution of \ion{Mg}{2} gas. For instance, high-resolution spectroscopy reveals the detailed kinematics of discrete absorbers around galaxies~\citep{Steidel2002, Kacprzak2010, Nielsen2015}. However, such observations typically require a resolving power of $R \gtrsim 15{,}000$, and are thus limited to small sample sizes~\citep{Vogt1994, Dekker2000, Green2012}. Alternatively, large-scale surveys provide abundant samples that reveal global kinematic trends~\citep{GC2011, Kauffmann2017, Huang2021}, yet they frequently suffer from low sensitivity to weak absorbers~\citep{ZM2013, Napolitano2023}. Additionally, recent studies have attempted to infer kinematic information from stacked spectra by approximating composite line profiles with simple parametric forms, such as exponential~\citep{Wild2008}, Lorentzian~\citep{Lundgren2009}, Gaussian~\citep{Chen2010a}, or two-component models~\citep[e.g.,][]{Huang2016, Anand2021, Chang2025}. One interesting finding from these studies is that the velocity dispersion of \ion{Mg}{2} absorbers is smaller than the typical virial dispersions of their dark matter halos~\citep{Zhu2014, Huang2016, Anand2021}.

However, further investigation of the cool cloud kinematics has been limited by the difficulty of moving beyond simple dispersion-based models, mainly due to the lack of a large absorber catalog in the redshift range of the LRGs. Overcoming this limitation is imperative given the complex origins of the CGM gas. In particular, the cool CGM reservoir could be clouds \emph{accreted} from the metal-polluted intergalactic medium (IGM)~\citep{Keres2005, OD2008, Oppenheimer2010}, radiatively cooled from hot outflows driven by past star formation or Active Galactic Nucleus~(AGN) feedback~\citep{Thompson2016, Sparre2019, GO2020}, condensed from the hot CGM due to turbulent mixing~\citep{McCourt2012, Sharma2012, Voit2017}, or simply associated with the CGM of satellite galaxies~\citep{TC2010, GC2011, Appleby2023}. A combination of these components is also likely, further complicating the kinematic origin of cool CGM gas. With a factor of $\sim 3.2$ increase in the number of quasar spectra compared to Sloan Digital Sky Survey~\citep[SDSS;][]{York2000} Data Release 16~\citep[DR16;][]{Ahumada2020}, DESI DR2 provides an unprecedented opportunity to develop new observational probes and sophisticated kinematic models that can elucidate the complex nature of the cool CGM around LRGs.

Recently,~\citet{Zu2021} (hereafter referred to as \defcitealias{Zu2021}{Zu21}\citetalias{Zu2021}) developed a theoretical framework for inferring cool cloud kinematics from the redshift-space cross-correlation function $\xsgc$ between galaxies and discrete \ion{Mg}{2} clouds at projected distance $r_p$. The method makes use of the convolution between the real-space galaxy-cloud correlation $\xrgc$ and the LOS velocity distribution of clouds $p(\vlos \mid r_p, \, y)$,
\begin{equation}
    \xsgc (r_\pi \mid r_p) + 1 = \left[ \xrgc \left(\sqrt{r^2_p + y^2} \right) + 1 \right] \ast p(\vlos \mid r_p, \, y) ,
    \label{eqn:r2s}
\end{equation}
where $y$ and $r_\pi$ are the LOS distances between galaxy-cloud pairs in real and redshift space, respectively~\citep{Peebles1980, Fisher1995}. Using the SDSS \ion{Mg}{2} absorber catalog, \citetalias{Zu2021} demonstrated that different models of $p(\vlos)$ predict distinct shapes of $\xsgc$, allowing the reconstruction of cool cloud kinematics from $\xrgc$. Intriguingly, they showed that an outflowing component ($\sim 250 \, \mathrm{km \, s^{-1}}$) at large radii ($\sim 400 \, h^{-1}\mathrm{kpc}$) is favored by the SDSS data. In this work, we improve upon the~\citetalias{Zu2021} framework in two key aspects, one observational and one theoretical, to extract the rich kinematic information from DESI DR2. In particular, we hope to detect any potential signatures of such an intriguing cool CGM outflow within the halos of DESI LRGs.

Observationally, the~\citetalias{Zu2021} method relies on identifying discrete \ion{Mg}{2} absorbers in quasar spectra, a process biased against the weak absorbers that dominate the population~\citep{Churchill1999, Nestor2005, Narayanan2007}. In addition, the absorber catalog may be contaminated by a small fraction of false positives due to random noise in the quasar continuum. In this work, we circumvent those issues by converting the stacked \ion{Mg}{2} absorption line profile into a $\xi^s(r_p, \, r_\pi)$ measurement, avoiding the need to detect discrete absorbers.

On the theoretical front, the kinematic models employed by~\citetalias{Zu2021} only consider radial velocities in their prediction of the LOS velocity distribution $p(\vlos)$. In this work, we turn to high-resolution hydrodynamical simulations for insights into a realistic 3D model of cool cloud kinematics~\citep{Muratov2015, Hopkins2018, Hummels2019}. As a first step toward building a realistic model for $p_\mathrm{los}$, we examine the cool cloud kinematics in the IllustrisTNG simulation due to its ideal combination of volume and resolution~\citep{Nelson2020, DeFelippis2021, Ramesh2023a, Ramesh2023b, Weng2024}, and extract a comprehensive analytic model from the 3D velocity distribution of the simulated gas clouds.

This paper is organized as follows. We describe the DESI DR2 data, including the LRG catalog, quasar continua, and an ancillary \ion{Mg}{2} absorber catalog in \S~\ref{sec:data}. In \S~\ref{sec:rsd} we present the stacked measurements of the equivalent width profiles of the \ion{Mg}{2} absorption and convert them into the RSD signals. In \S~\ref{sec:tng}, we develop a 2D velocity distribution model for cool gas by tracking the dynamical histories of CGM gas cells in the TNG50 simulation. We then explore potential evidence for cool outflows using the distinctive Finger-of-God~\citep[FoG;][]{Jackson1972} effect in the stacked RSD measurement in \S~\ref{sec:fog} and discuss the physical implications. We summarize our findings and look to the future in \S~\ref{sec:summary}.

\section{Data}
\label{sec:data}

Our study makes use of the data products from the DESI survey, a Stage~IV dark-energy survey conducted with the 4-m Nicholas U. Mayall Telescope at Kitt Peak~\citep{DESI2022}. DESI features a robotic, fiber-fed, highly multiplexed spectroscopic facility with 5000 independently positioned fibers over a $3.2^\circ$ diameter field of view, enabling simultaneous spectroscopy of nearly 5000 targets per pointing~\citep{DESI2016a, DESI2016b, Miller2024, Poppett2024}. The eight-year survey covers $\sim 17{,}000 \, \mathrm{deg^2}$~\citep{Schlafly2023} and is expected to deliver 63 million spectra of galaxies and quasars~\citep{Guy2023}. The upcoming DESI DR2 spans the first three years of the survey (2021 May 14 to 2024 April 9) and is scheduled for public release in 2027. For our analysis, we use the LRG and quasar samples from the $\texttt{loa}$ run of the DESI DR2~\citep{Karim2025a, Karim2025b}.

\subsection{Luminous Red Galaxies}
\label{subsec:lrg}

The target selection for DESI LRGs is primarily based on the imaging data from the Legacy Surveys, using a set of color and magnitude cuts from~\citealt{Zhou2020}. Their redshifts are measured from DESI spectra using the template-fitting classifier \textsc{Redrock}\footnote{\url{https://github.com/desihub/redrock}}~\citep{Brodzeller2023, Anand2024, Bailey2026}. For our analysis, we use the LRG sample from the DESI Large-Scale Structure (LSS) catalog, which includes veto masks tailored for clustering studies. These masks retain $\simeq 98.9\%$ of the observed LRG targets. To ensure high-quality redshifts for RSD measurements, we apply three additional quality cuts: $\texttt{ZWARN}=0$, $\texttt{ZERR}\leq 10^{-4}$, and $\Delta \chi^2 > 15$~\citep{DESI2024}. These cuts yield $N_\mathrm{LRG} = 2{,}414{,}573$ LRGs over $0.25 < z < 1.25$, with a median redshift of $z \simeq 0.61$. For the galaxy-absorber cross-correlation, we use the accompanying random LRG catalog from the LSS directory, which is roughly 120 times as large as the data catalog.

\subsection{Quasar Spectra and Continuum Determination}
\label{subsec:quasar}

We adopt the quasar catalog from DESI DR2, including $2{,}456{,}178$ quasars over a redshift range of $0.05 < z < 5.80$. DESI first identifies quasar candidates using the generic \textsc{Redrock} classifier, then applies two quasar-specific algorithms: broad \ion{Mg}{2} emission and the convolutional neural network \textsc{QuasarNET}~\citep{BB2018, Farr2020}. These two algorithms improve the quasar sample completeness by $\sim 10\%$~\citep{Chaussidon2023}, mostly due to reclassifying the \textsc{Redrock}-identified galaxies with broad \ion{Mg}{2} emission into quasars. In addition, the afterburners recover a population of misclassified quasars at redshifts lower than $z = 1.7$, where \textsc{Redrock} is limited by the spectral template range available for quasars~\citep{Alexander2023}. Since our stacking analysis is insensitive to the detailed spectral class provided the redshift is secure, we apply only $\texttt{ZWARN}=0$ to exclude spectra with unreliable redshifts, yielding $2{,}398{,}161$ quasars over redshift $0.05 < z < 4.80$. Additional quality control for noisy spectra is deferred to the stacking pipeline, as detailed in \S~\ref{subsec:stack}.

Our stacked RSD measurement depends on accurately determining the continuum flux $\Fcon(\lambda)$ for each individual quasar spectrum $\Fqso(\lambda)$. We estimate continua with the Non-negative Matrix Factorization~\citep[NMF;][]{LS1999} algorithm, adapted for spectroscopic studies by~\citet{ZM2013}. Using the NMF implementation\footnote{\url{https://github.com/guangtunbenzhu/NonnegMFPy}} of~\citet{Zhu2016} to model and subtract prominent spectral features, we obtain an estimate of $\Fcon(\lambda)$ and derive the \emph{normalized} spectrum for each quasar as
\begin{equation}
    \fqso(\lambda) \equiv \frac{\Fqso(\lambda)}{\Fcon(\lambda)} ,
    \label{eqn:norm_spec}
\end{equation}
which will later serve as the input to our spectral stacking. However, NMF continuum fitting cannot perfectly trace the intrinsic continuum of every quasar spectrum. After stacking, we find that systematic continuum residuals persist well beyond the absorption wavelength window. We resolve this issue by constructing a randomly shuffled quasar stack and applying a linear fit to the random normalized composite continuum while masking the absorption window (Appendix~\ref{app:stk_cor}).

\subsection{Ancillary Catalog of \texorpdfstring{\ion{Mg}{2}}{Mg II} Absorbers}
\label{subsec:mg2_cat}

Our fiducial analysis uses the \ion{Mg}{2} RSD measurement from spectral stacking, a technique that does not require detecting individual absorbers. For the sake of comparison, however, we also compute galaxy-\ion{Mg}{2} cross-correlations using a catalog of individually detected DESI \ion{Mg}{2} absorbers. This ancillary \ion{Mg}{2} absorber catalog is an internal value-added product to be released with DESI DR2, superseding its DR1 predecessor\footnote{\url{https://data.desi.lbl.gov/doc/releases/dr1/vac/mgii-absorber/}}. Absorbers are identified in quasar spectra of the same afterburner catalog as our stacking sample by an autonomous pipeline, and we refer interested readers to~\citet{Napolitano2023} for technical details. The absorber catalog contains $187{,}641$ discrete \ion{Mg}{2} systems over $0.28 < z < 1.30$, with an estimated purity exceeding $99\%$.

For clustering measurements from these discrete absorbers, we construct a random \ion{Mg}{2} absorber catalog $100$ times as large as the data using the method introduced by~\citetalias{Zu2021}. Instead of generating random angular coordinates for the random absorbers, we draw a random sample of angular positions from the \emph{parent} quasar catalog (including sightlines without absorber detections). This procedure guarantees that the random catalog inherits the correct observational weights of the observed quasars. We then assign them absorber redshifts randomly drawn from the data absorber catalog, while requiring each random redshift to fall within the search window of its assigned quasar.

We compute $\xsga (r_p, \, r_\pi)$ for the discrete galaxy-absorber\footnote{We distinguish between stacked and discrete galaxy-\ion{Mg}{2} cross-correlation measurements using superscripts (subscripts) $\mathrm{gc}$ (galaxy-cloud) and $\mathrm{ga}$ (galaxy-absorber), respectively.} sample using the~\citet{LS1993} estimator,
\begin{equation}
    \xsga(r_p, \, r_\pi) = \frac{D_g D_a - D_g R_a - R_g D_a + R_g R_a}{R_g R_a} ,
    \label{eqn:ls}
\end{equation}
where $D_gD_a$, $D_gR_a$, $R_gD_a$, and $R_gR_a$ are the normalized galaxy-absorber data-data, data-random, random-data, and random-random pair counts within bins of $(r_p, \, r_\pi)$. Both pairwise-inverse-probability weights and angular up-weighting are applied in our measurements~\citep{BP2017, PB2017}. We then compute the projected galaxy-absorber cross-correlation $\wpga (r_p)$ by integrating $\xsga$ along the LOS:
\begin{equation}
    \wpga(r_p) = \int_{-r_{\pi, \max}}^{+r_{\pi, \max}} \xsga(r_p, \, r_\pi) \, \mathrm d r_\pi,
\end{equation}
with an integration limit of $r_{\pi, \max} = 40 \, h^{-1} \mathrm{Mpc}$. The uncertainty matrices for both $\xsga$ and $\wpga$ are estimated via Jackknife resampling over $200$ approximately equal-area regions of the DESI DR2 footprint. For the rest of the paper, we refer to all the measurements based on this ancillary absorber catalog as the ``discrete'' measurements.

\subsection{TNG50 Simulation}

To gain theoretical insights on the dynamical CGM, we make use of IllustrisTNG\footnote{\url{https://www.tng-project.org}}~\citep{Pillepich2018, Nelson2018, Naiman2018, Marinacci2018, Springel2018}, a set of large magnetohydrodynamical simulations run with the moving mesh code \textsc{Arepo}~\citep{Springel2010} under Planck cosmology~\citep{Planck2016}. In particular, we focus on the highest-resolution run of the program, the TNG50-1 simulation~\citep{Nelson2018, Pillepich2018}, which evolves a $(51.7 \, \mathrm{cMpc})^3$ volume with $2160^3$ resolution elements (different but analogous to ``particles''). The baryonic and dark matter mass resolutions are $8.5 \times 10^4 \, \mathrm{M_\odot}$ and $4.5 \times 10^5 \, \mathrm{M_\odot}$, respectively. More importantly, while the adaptive gas resolution reaches typical cell sizes of $\sim 100$--$200 \, \mathrm{pc}$ in dense regions, the simulation can still resolve the diffuse CGM on $\mathrm{kpc}$ scales~\citep{Nelson2020}. This combination of high spatial resolution and full baryonic physics allows us to investigate realistic CGM kinematics in LRG-like galaxies inside the cosmic web.

\section{Stacked \texorpdfstring{\ion{Mg}{2}}{Mg II} RSD Measurement}
\label{sec:rsd}

As shown in Equation~\eqref{eqn:r2s}, inferring $p(\vlos)$ requires deconvolving the redshift-space galaxy-cloud cross-correlation $\xsgc$ from the 3D real-space correlation $\xrgc$. In order to perform this deconvolution, we directly measure $\xsgc$ using spectral stacking, while deriving $\xrgc$ from the projected galaxy-cloud cross-correlation function $\wpgc$.

Briefly, we first develop a spectral-stacking pipeline for DESI DR2 and measure the average \ion{Mg}{2} absorption line profiles (normalized by the continuum), $\langle \fMg (\lamr \mid r_p)\rangle$, across various projected radii $r_p$ around LRGs. Next, we integrate the stacked line profiles over $\lamr$ to obtain the mean \ion{Mg}{2} rest-frame equivalent width (REW) as a function of $r_p$, and convert it to the projected galaxy-cloud cross-correlation $\wpgc(r_p)$. We then deproject $\wpgc(r_p)$ to obtain the 3D real-space cross-correlation $\xrgc(r)$. Finally, we convert the stacked line profiles into the 2D redshift-space correlation function $\xsgc(r_p, \, r_\pi)$. We describe these stacking-based measurements in detail below, but the impatient readers can skip to \S~\ref{subsec:xi_s} for the final $\xsgc$ measurements.

\subsection{Composite Absorption Profile from Spectral Stacking}
\label{subsec:stack}

\begin{figure*}
    \includegraphics[width=\linewidth]{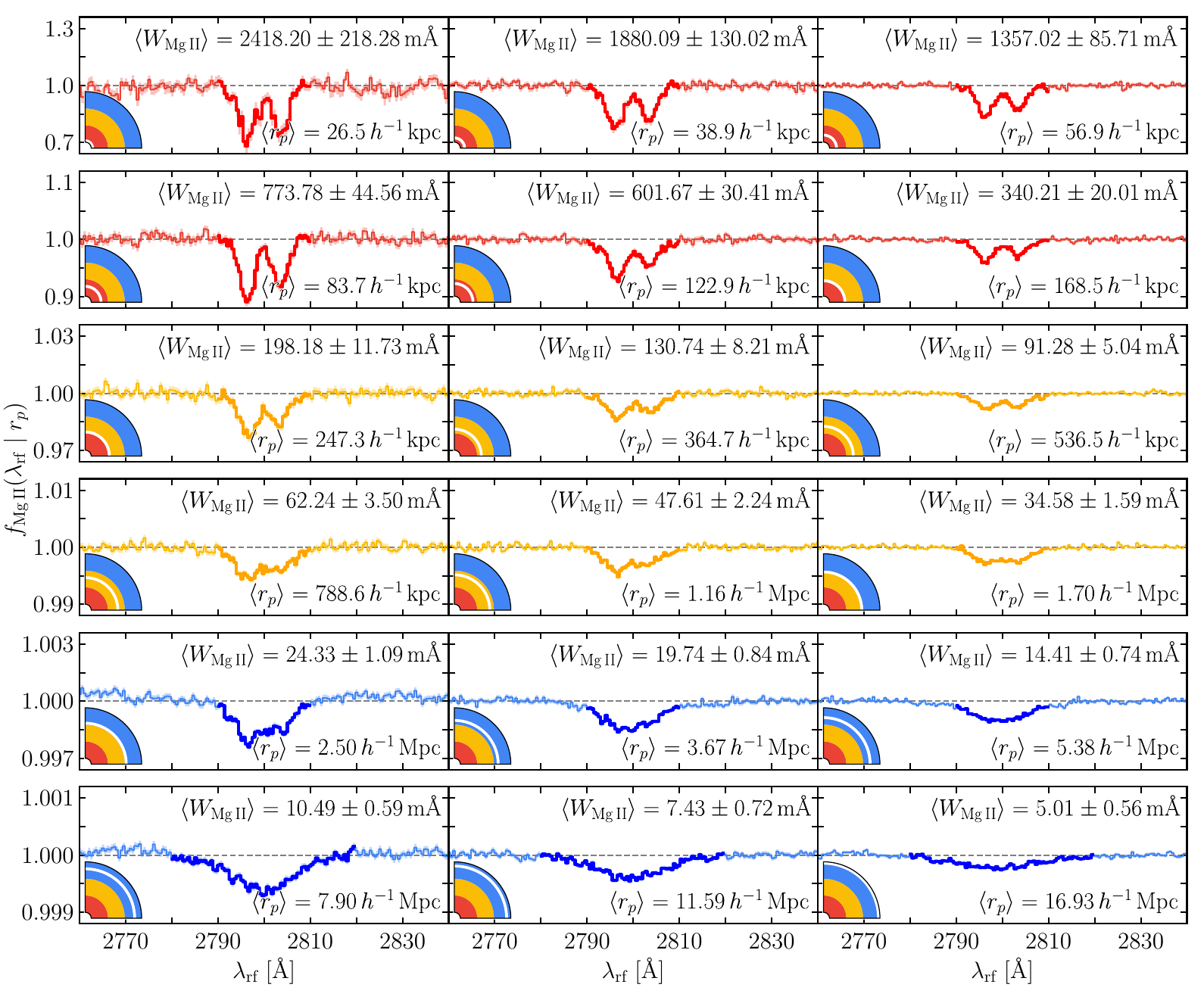}
    \caption{\textit{Stacked \ion{Mg}{2} absorption line profiles across 18 bins of projected distance $r_p$.} In each panel, the colored solid curve is the fractional absorption by \ion{Mg}{2} as a function of rest-frame wavelength at that average distance $\langle r_p \rangle$ (listed in the bottom right corner). The shaded band around the curve indicates the $1\sigma$ uncertainties derived from bootstrapping, and the thicker segment of the curve indicates the integration window for calculating the REW $\langle \WMg \rangle$ (listed in the top right corner). In the bottom left corner, a logarithmic pie chart indicates the corresponding $r_p$ bin (white wedge) relative to the full $r_p$ range, with the colors of the three ring areas matched to that of the solid curves. Note that the $y$-axis limits of the 18 panels differ between rows but remain consistent within a given row.}
    \label{fig:stack}
\end{figure*}

\begin{figure}
    \includegraphics[width=\columnwidth]{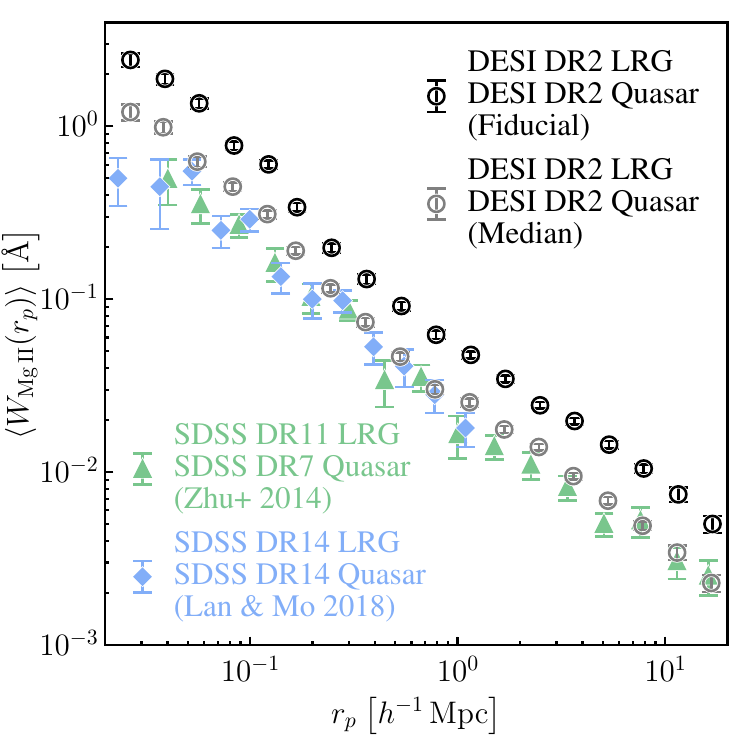}
    \caption{\textit{Rest-frame equivalent widths of \ion{Mg}{2} absorption as functions of projected distance away from the LRGs.} Black and gray circles denote our geometric mean and median stacked measurements from DESI DR2, while green filled triangles and blue filled diamonds represent the stacked results from~\citealt{Zhu2014} and~\citealt{LM2018}, respectively. To facilitate a direct comparison, we convert their original measurement from physical to comoving distances using the mean redshift of their respective LRG samples.}
    \label{fig:rew_rp}
\end{figure}

To measure the composite \ion{Mg}{2} absorption line profile $\langle \fMg (\lamr \mid r_p) \rangle$ within different comoving $r_p$ bins around LRGs, we select all LRG-quasar pairs with transverse separations below $20 \, h^{-1} \mathrm{Mpc}$ and redshift differences above $0.1$. To achieve a clean stacking measurement, we trim each quasar spectrum according to the same criteria adopted in~\citealt{Napolitano2023}. For a quasar at redshift $z_q$, we restrict the spectrum to wavelengths redward of the quasar Ly$\alpha$ emission to avoid the Ly$\alpha$ forest, and $\Delta z = 0.03$ blueward of the quasar \ion{Mg}{2} emission line. Therefore, the minimum and maximum wavelengths of the wavelength range are $\lambda_\mathrm{min} = 1216 (1 + z_q) \, \textup{\AA}$ and $\lambda_\mathrm{max} = 2800 (0.97 + z_q) \, \textup{\AA}$, respectively.

For an LRG with redshift $z_g$, we shift the normalized and trimmed spectra $\fqso(\lambda \in \left[ \lambda_\mathrm{min}, \, \lambda_\mathrm{max} \right])$ of all neighboring quasars into the rest frame at $z_g$, $\fqso(\lamr \equiv \lambda / (1 + z_g))$, and interpolate them onto a common wavelength grid matched to the native DESI wavelength resolution at the effective\footnote{While the full LRG sample has a median redshift of $z = 0.61$, the paired LRGs have a median of $z = 0.58$. We therefore adopt $\zeff = 0.58$ as the effective redshift for the following analysis.} redshift of $\zeff = 0.58$. We then stack the individual absorption signals of all LRGs in annuli of fixed physical radii, and convert the measurements to comoving coordinates using $\zeff$. This produces a preliminary stacked measurement of the absorption line profile at each $r_p$, $\langle \flin (\lamr \mid r_p) \rangle$. During the stacking, we first apply a $5 \sigma$-clipping to remove abnormal flux values and then measure the \emph{geometric} mean flux at each wavelength pixel. Since absorption is a multiplicative effect, calculating the geometric mean of the spectra provides a natural estimate of the arithmetic mean of the optical depths.

We develop a two-step correction method to determine the residual continuum in $\langle \flin(\lamr \mid r_p) \rangle$. First, we build a random quasar sample ten times as large as the data sample and process it through the identical stacking pipeline, producing a composite ``line'' spectrum that only includes the continuum, $\langle \clin(\lamr) \rangle$. Second, we normalize the data composite by the random composite to obtain
\begin{equation}
    f^\prime_\mathrm{line}(\lamr \mid r_p) \equiv \frac{\langle \flin(\lamr \mid r_p) \rangle}{\langle \clin(\lamr) \rangle} ,
\end{equation}
which we then fit with a linear function $f^\prime_\mathrm{linear}(\lamr \mid r_p)$, while masking out the wavelength range from $2780$ to $2820 \, \textup{\AA}$. We describe the technical details of this two-step procedure in Appendix~\ref{app:stk_cor}.

Finally, we compute our fiducial stacked \ion{Mg}{2} absorption line profile as
\begin{equation}
    \fMg(\lamr \mid r_p) = \frac{f^\prime_\mathrm{line}(\lamr \mid r_p)}{f^\prime_\mathrm{linear}(\lamr \mid r_p)} ,
    \label{eqn:fmg}
\end{equation}
and the corresponding REW as
\begin{equation}
    \langle \WMg (r_p) \rangle = \int \left[ 1 - \fMg (\lamr \mid r_p) \right] \, \mathrm d \lamr .
    \label{eqn:Wmg}
\end{equation}
The measurement uncertainties of the composite profiles and REWs are calculated using $200$ bootstrap realizations, resampling LRG--quasar pairs across sky subregions. We note that some previous studies~\citep{Zhu2014, LM2018} fit double-Gaussian profiles to measure REWs, whereas we integrate the composite spectrum directly~\citep{Wu2024}. As indicated in our later redshift-space analysis, a double-Gaussian fails to capture the broad line wings in redshift space and can potentially underestimate the REWs.

Figure~\ref{fig:stack} presents our final stacked absorption line profiles in all $18$ $r_p$ bins across three orders of magnitude in distance scales, ranging from $\sim 20 \, h^{-1} \mathrm{kpc}$ (top left) to $\sim 20 \, h^{-1} \mathrm{Mpc}$ (bottom right). Note that the $y$-axis ranges are the same in each row, but decrease from $\pm 30 \%$ in the top row to $\pm 0.1 \%$ in the bottom row. In each panel, the solid curve with a $1\sigma$ error band shows the final composite of $\fMg (\lamr \mid r_p)$, with the thicker segment highlighting the integration windows for computing $\langle \WMg (r_p) \rangle$. The $18$ profiles are color-coded according to the average radius $\langle r_p \rangle$ of the bin, indicated by the white stripe inside the tri-colored quadrant in the bottom left corner ($20 \, h^{-1} \mathrm{kpc} \leq \langle r_p \rangle < 200 \, h^{-1} \mathrm{kpc}$: red, $200 \, h^{-1} \mathrm{kpc} \leq \langle r_p \rangle < 2 \, h^{-1} \mathrm{Mpc}$: yellow, $2 \, h^{-1} \mathrm{Mpc} \leq \langle r_p \rangle < 20 \, h^{-1} \mathrm{Mpc}$: blue). Although the average REW of the absorption decreases from $1839 \, \textup{m\AA}$ ($\langle r_p \rangle = 26.5 \, h^{-1} \mathrm{kpc}$) to $4.38 \, \textup{m\AA}$ ($\langle r_p \rangle = 18.14 \, h^{-1} \mathrm{Mpc}$), the fractional error in the REW measurement stays roughly constant at the 5--10$\%$ level across all radial bins.

We compare our DESI REW profiles around the LRGs $\langle \WMg (r_p) \rangle$ (open symbols) with the previous SDSS measurements (solid symbols) in Figure~\ref{fig:rew_rp}. Black open circles with errorbars indicate our fiducial DESI measurement using the geometric mean stack. The REW profile clearly deviates from a single power-law shape, exhibiting a steeper rise within $500 \, h^{-1} \mathrm{kpc}$ than on large scales. This slope change is reminiscent of the physical transition between the one-halo and two-halo regimes, often seen in the correlation functions of galaxies. Therefore, from this characteristic transition we anticipate that the typical radius of DESI LRG halos is roughly $500 \, h^{-1} \mathrm{kpc}$, corresponding to a halo mass of roughly $1.5 \times 10^{13} \, \mathrm{M_\odot}$ at $\zeff = 0.58$.

From the same DESI DR2 data, we measure another stacked REW profile using the median stacked spectra instead of the geometric mean used for the fiducial profile. This alternative DESI REW profile (gray open circles with errorbars) has an amplitude that is lower by $\sim 66 \%$ compared to that of the fiducial profile. For comparison, light blue diamonds and green triangles are the stacked REW profiles measured from the SDSS DR14~\citep{LM2018} and SDSS DR7~\citep{Zhu2014}, respectively. They are both consistent with the DESI median profile, and are therefore lower by roughly a factor of two than our DESI fiducial profile. Apart from the amplitude difference, the DESI REW profiles exhibit a much clearer one-halo to two-halo transition than the SDSS results, due to the significantly improved measurement uncertainties in DESI (by a factor of three).

\subsection{\texorpdfstring{Projected Correlation $w^{\mathrm{gc}}_p$ from Equivalent Width Profile $\langle W_\mathrm{Mg \, \textup{\textsc{ii}}} \rangle$}{Projected Correlation wp from Equivalent Width W}}
\label{subsec:wp}

\begin{figure}
    \includegraphics[width=\columnwidth]{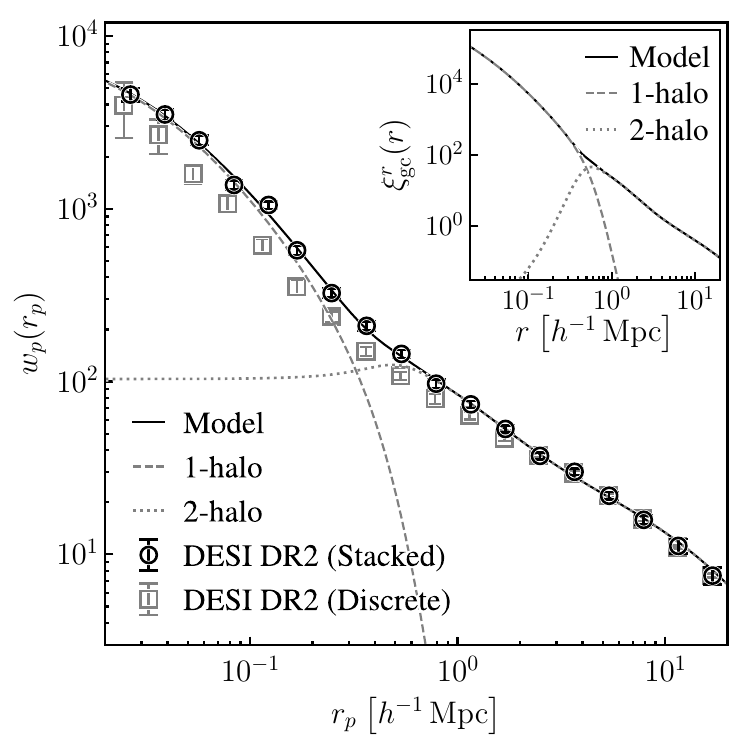}
    \caption{\textit{Projected LRG-cloud cross-correlation function $w_p(r_p)$.} Black open circles with errorbars indicate the $\wpgc(r_p)$ derived from our DESI stacked REW profile, while the black solid curve is the prediction by our best-fitting model, comprising the 1-halo (gray dashed) and 2-halo (gray dotted) contributions. The inset panel shows the corresponding 1-halo and 2-halo contributions to the best-fitting 3D real-space correlation functions $\xrgc$ (black solid). For reference, the gray open squares denote $\wpga(r_p)$ measured from the ancillary catalog of discrete absorbers.}
\label{fig:wp_rp}
\end{figure}

We now derive the 2D projected galaxy-cloud cross-correlation $\wpgc (r_p)$, which can then be deprojected into the 3D real-space galaxy-cloud cross-correlation $\xrgc$ required for the deconvolution in Equation~\eqref{eqn:r2s}. To obtain $\wpgc (r_p)$, we reinterpret the REW profile $\langle \WMg (r_p) \rangle$ as an excess \ion{Mg}{2} column density profile $\langle \NMg (r_p) \rangle$ and then convert it into $\wpgc (r_p)$.

Prior works~\citep[e.g.,][]{Zhu2014, Wu2024} demonstrated that the REW profile can be mapped to column density with an explicit treatment of saturation. We follow~\citet{Zhu2014} and apply a saturation correction based on the doublet ratio, defined as
\begin{equation}
    \fdrt \equiv \frac{\WMg^{\lambda2796}}{\WMg^{\lambda2803}} ,
\end{equation}
while assuming the $\lambda2803$ component remains optically thin for all absorbing clouds. This doublet ratio is a function of the total REW of the doublet $\langle \WMg \rangle$. We combine the DESI DR2 \ion{Mg}{2} absorber catalog ($\WMg > 0.05 \, \textup{\AA}$) with a sample of weak absorbers from~\citealt{Churchill1999} and~\citealt{Narayanan2007} ($0.018 < \WMg < 0.3 \, \textup{\AA}$) to determine $\fdrt \big( \langle \WMg \rangle \big)$. The value of $\fdrt$ is bound between $1.75$ in the optically-thin limit and unity in the saturation limit, with an REW dependence that can be well described by
\begin{equation}
    \frac{\log_{10} \fdrt}{\log_{10} 1.75} = \frac{1}{2} \mathrm{erfc}\!\left[\frac{\log_{10}\big( \langle \WMg \rangle / W_0 \big)}{\sqrt{2} \, \sigma}\right] ,
    \label{eqn:dr}
\end{equation}
with $\sigma = 0.92$ and $W_0 = 0.82 \, \textup{\AA}$ (See Appendix~\ref{app:sat_cor}).

The intrinsic (i.e., in the optically-thin limit) REW contribution from the $\lambda2803$ component is therefore
\begin{equation}
    \langle W^{\lambda 2803}_\mathrm{Mg \, \textsc{ii}} \rangle = \frac{\langle \WMg \rangle}{1 + \fdrt \big(\langle \WMg \rangle \big)} ,
\end{equation}
which can be converted to an excess ionic column density using the linear curve-of-growth relation~\citep{Spitzer1978},
\begin{equation}
    \big\langle \NMg (r_p) \big\rangle = \frac{m_e c^2}{\pi e^2} \frac{\langle W^{\lambda2803}_\mathrm{Mg \, \textsc{ii}} \rangle}{f_0 \, \lambda^2_0} ,
\end{equation}
where $\lambda_0 = 2803.5315 \, \textup{\AA}$ is the rest-frame wavelength of the $\lambda2803$ component and $f_0 = 0.3058$ is its corresponding oscillator strength~\citep{Morton2003}. We note that a more sophisticated correction using the \ion{Si}{2} line was employed by~\citet{Wu2024} for the stacked profiles around the emission-line galaxies (ELGs). Nonetheless, we choose to apply the simpler $\fdrt$-based correction as it is adequate for the LRGs that generally exhibit much weaker \ion{Mg}{2} absorptions than the ELGs~\citep{LM2018}.

The projected cross-correlation function is tied to the excess column density profile via
\begin{equation}
    w^{\mathrm{gc}}_p(r_p) = \frac{\langle \NMg (r_p) \rangle}{\nMg (\zeff)} ,
\end{equation}
where $\nMg (\zeff)$ is the average number density of \ion{Mg}{2} ions in the Universe at $\zeff$. We adopt the empirical method proposed by \citetalias{Zu2021} and infer $\nMg (\zeff) = 3.64 \times 10^{-15} \, \mathrm{cm^{-3}}$ by matching $\langle \NMg (r_p) \rangle$ to the projected cross-correlation $w^\mathrm{ga}_p$ between the LRGs and discrete absorbers at $r_p \gtrsim 2 \, h^{-1} \mathrm{Mpc}$ (gray squares in Figure~\ref{fig:wp_rp}). Consequently, the final conversion from our REW profile to the projected correlation is
\begin{equation}
    \wpgc(r_p) = 4.49 \times 10^{3} \, h^{-1} \mathrm{Mpc} \times \frac{\langle W^{\lambda2803}_\mathrm{Mg \, \textsc{ii}} (r_p) \rangle}{1 \, \textup{\AA}} ,
\end{equation}
The result is shown as the black circles with errorbars in Figure~\ref{fig:wp_rp}. Compared with the discrete measurement $\wpga$ (gray squares), the stacked measurement $\wpgc$ is higher by $\gtrsim 30 \%$ on scales below $\sim 1 \, h^{-1} \mathrm{Mpc}$, i.e., the 1-halo regime. This discrepancy suggests that roughly half of the \ion{Mg}{2}-bearing gas may be missing from the discrete catalog.

\subsection{3D Real-space Correlation \texorpdfstring{$\xrgc$}{xi} from deprojecting \texorpdfstring{$\wpgc$}{wp}}
\label{subsec:xi_r}

Following~\citetalias{Zu2021}, we deprojected the stacked measurement of $\wpgc(r_p)$ into the 3D isotropic real-space correlation $\xrgc(r)$ using a Bayesian forward-modeling approach. In particular, we predict $\wpgc(r_p)$ from $\xrgc (r)$ using
\begin{equation}
    \wpgc(r_p) = \int_{-\infty}^{+\infty} \xrgc (\sqrt{r_p^2 + r_\pi^2}) \, \mathrm d r_\pi ,
    \label{eqn:wpgc}
\end{equation}
We adopt an analytic model of $\xrgc(r)$ consisting of two components,
\begin{equation}
    \xrgc(r) = \xrOh(r) + \xrTh(r) ,
    \label{eqn:xi_1h2h}
\end{equation}
where $\xrOh$ and $\xrTh$ are the 1-halo and 2-halo terms, respectively. The 1-halo term can be written as:
\begin{equation}
    \xrOh(r) = \frac{\rhcl(r)}{\bar \rho_\mathrm{cool}} ,
    \label{eqn:xi_1h}
\end{equation}
where $\rhcl(r)$ is the average density profile of cool, \ion{Mg}{2}-enriched gas within the dark matter halos and $\bar \rho_\mathrm{cool}$ is the mean density of such gas in the whole Universe.

Assuming the cool gas roughly follows the dark matter distribution within halos, we have
\begin{equation}
    \rhcl(r) = \fcgm \cdot \rho_\mathrm{NFW}(r \mid \Mvir, \, c_h) \cdot \ftrn(r) ,
\end{equation}
where $\fcgm$ is the cool gas-to-halo mass ratio within the halo, $\rho_\mathrm{NFW} (r \mid \Mvir, \, c_h)$ is the Navarro-Frenk-White~\citep[NFW;][]{NFW1997} density profile for a halo with virial mass $M_\mathrm{vir}$ and concentration $c_h$, and $\ftrn$ accounts for the splashback transition near the halo virial radius $\rvir$,
\begin{equation}
    \ftrn = \left[ 1 + \left( \frac{r}{1.495 \, \rvir} \right)^4 \right]^{-2} .
\end{equation}

Defining the cool gas-to-halo mass ratio of the Universe as
\begin{equation}
    \bar f_\mathrm{cool} = \frac{\bar \rho_\mathrm{cool}}{\rho_m} ,
\end{equation}
where $\rho_m$ is the mean matter density of the Universe, we rewrite the 1-halo term as
\begin{equation}
    \xrOh(r) = b_\mathrm{1h} \frac{\rho_\mathrm{NFW}}{\rho_m} \ftrn (r) ,
\end{equation}
where $b_\mathrm{1h} = \fcgm / \bar f_\mathrm{cool}$ is the mass bias of cool gas within the halos. Since the cool, enriched mass is primarily found in the CGM, we expect $b_\mathrm{1h}$ to be significantly above unity.

We adopt the same model for the 2-halo term as in~\citetalias{Zu2021}. The model assumes that the large-scale clustering of cool, enriched gas around LRGs is
\begin{equation}
    \xrTh(r) = b_\mathrm{2h} \cdot \xi_\mathrm{mm} (r) \cdot f_\mathrm{ex} (r) ,
    \label{eqn:xi_2h}
\end{equation}
where $b_\mathrm{2h}$ is the product of $b_\mathrm{LRG}$ and $b_\mathrm{cool}$ (i.e., the linear biases of LRGs and cool gas, respectively), $\xi_\mathrm{mm} (r)$ is the non-linear matter auto-correlation, and $f_\mathrm{ex} (r)$ describes the halo exclusion effect when approaching the 1-halo regime
\begin{equation}
    f_\mathrm{ex} (r) = \frac{1}{2} \left[ 1 + \mathrm{erf} \left( \frac{r - \beta \, \rvir}{\sigma_\mathrm{ex}} \right) \right] .
\end{equation}

\begin{deluxetable*}{cccccc}
    \setlength{\tabcolsep}{2.0pt}
    \tablecaption{Best-fitting parameters for spatial-clustering two-component model. Values are maximum-likelihood estimates.
    \label{tab:wp_fit}}
    \tablehead{\colhead{$\log_{10} \Mvir$} & \colhead{$b_\mathrm{1h}$} & \colhead{$b_\mathrm{2h}$} & \colhead{$c_h$} & \colhead{$\beta$} & \colhead{$\sigma_\mathrm{ex}$}}
    \startdata
    $13.07^{+0.16}_{-0.13}$ & $3.42^{+0.97}_{-0.91}$ & $1.23^{+0.02}_{-0.02}$ & $7.17^{+1.39}_{-1.20}$ & $1.09^{+0.09}_{-0.06}$ & $0.15^{+0.04}_{-0.03}$ \\
    \enddata
\end{deluxetable*}

By combining Equations~\eqref{eqn:xi_1h2h} and \eqref{eqn:xi_1h}, we predict $\wpgc(r_p)$ using Equation~\eqref{eqn:wpgc} and fit to our stacked $\wpgc(r_p)$ measurement. We list the best-fitting parameters in Table~\ref{tab:wp_fit} and present the fitting results in Figure~\ref{fig:wp_rp}. The best-fitting model prediction (black solid curve) provides an excellent description of the stacked measurement (circles with errorbars), with the 1-halo and 2-halo components indicated by the gray dashed and dotted curves, respectively. The underlying best-fitting models of $\xrOh$ and $\xrTh$ are indicated by the gray dashed and dotted curves in the inset panel. The average virial radius of the LRG halos is $\rvir = 430^{+69}_{-41} \, h^{-1} \mathrm{kpc}$, consistent with our expectation from the 1-to-2 halo transition in the REW profile. Our inferred log-halo mass is $\log_{10} M_\mathrm{vir} = 13.07^{+0.16}_{-0.13}$, slightly lower than the constraint from the Halo Occupation Distribution (HOD) analysis~\citep[$13.40^{+0.02}_{-0.02}$][]{Yuan2024}.

\subsection{Stacked Measurement of \texorpdfstring{$\xsgc(r_p, \, r_\pi)$}{xi}}
\label{subsec:xi_s}

\begin{figure*}
    \includegraphics[width=\linewidth]{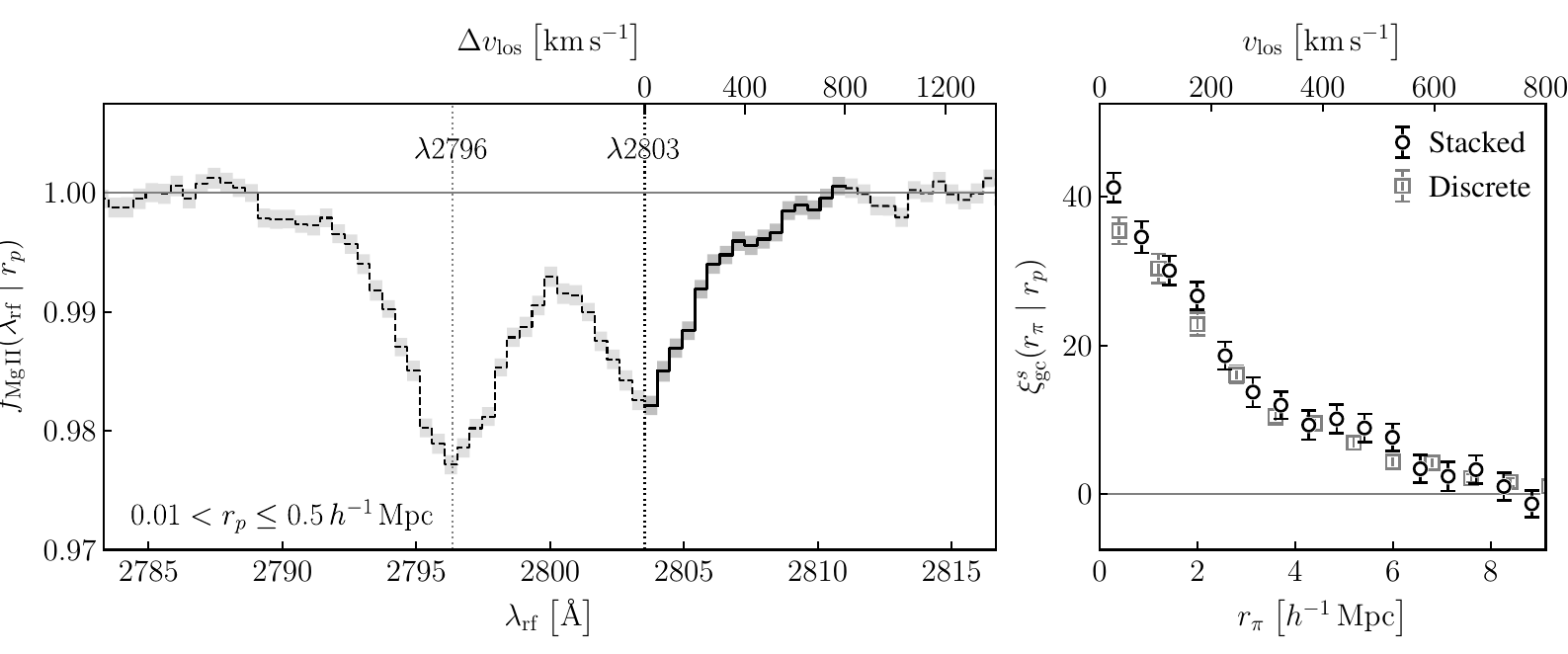}
    \caption{\textit{Conversion from the stacked \ion{Mg}{2} absorption profiles (}$\fMg(\lamr \mid r_p)$\textit{; left panels) to the redshift-space LRG-cloud cross-correlation functions ($\xsgc(r_\pi \mid r_p)$; right panel).} \textbf{Left:} Dashed gray histogram with shaded uncertainty bands is the stacked \ion{Mg}{2} absorption profiles measured at $0.01 < r_p \leq 0.5 \, h^{-1} \mathrm{Mpc}$. The highlighted black histogram is used for converting to our fiducial measurement of $\xsgc (r_\pi \mid r_p)$. \textbf{Right:} Open black circles with errorbars denote our fiducial stacked RSD measurement for the \ion{Mg}{2} gas within the LRG halos, derived from the $\lambda2803$ absorption profile. For reference, gray open squares are measured from the ancillary catalog of discrete absorbers. The stacked $\lambda$ RSD signal reveals a ``knuckle'' around $r_\pi = 5 \, h^{-1} \mathrm{Mpc}$ superimposed on the usual FoG effect.}
    \label{fig:f2xi}
\end{figure*}

\citetalias{Zu2021} measured the $\xi^s$ signal by cross-correlating the positions of galaxies with individual \ion{Mg}{2} absorbers, assuming that the cool CGM gas consists of discrete absorbers detected with well-defined redshifts. However, CGM gas may exist in the form of a continuous distribution of small cloudlets that do not have a well-defined shape or redshift~\citep[e.g.,][]{McCourt2018, Gronke2022}. In this scenario, the kinematic information is encoded in the shape of the absorption line profile rather than the line center (i.e., redshift), rendering the interpretation of $\xi^s$ ambiguous. Additionally, discrete measurements suffer from the incompleteness of weak absorbers inherent to all absorber-finding algorithms. In contrast, spectral stacking offers a natural mapping of the 2D redshift-space correlation function for a continuous distribution, because the stacked absorption profile preserves the redshift distortion of all the absorbing material along the LOS.

Therefore, we develop a novel pipeline for measuring $\xsgc(r_p, \, r_\pi)$ from the stacked \ion{Mg}{2} absorption line profile. This can be achieved by writing $\xsgc$ in the differential form
\begin{equation}
    \begin{aligned}
        \xsgc(r_p, \, r_\pi) & = \frac{\partial}{\partial r_\pi} \wpgc(r_p) \\
        & = \frac{m_e c^2}{\pi e^2} \frac{1}{f_0 \lambda^2_0 \nMg(\zeff)} \frac{\partial}{\partial r_\pi} \langle \WMg^\mathrm{\lambda2803} \rangle \\
        & = \frac{m_e c H(\zeff)}{\pi e^2 (1 + \zeff)} \frac{1 - \fMg (\lamr \mid r_p)}{f_0 \lambda_0 \nMg(\zeff)},
    \end{aligned}
    \label{eqn:xi^s}
\end{equation}
The final equality thus provides the direct conversion from $\fMg$ to $\xsgc$.

However, because Equation~\eqref{eqn:xi^s} assumes optically thin absorption, it is applicable only to the weaker $\lambda2803$ component. The stronger $\lambda2796$ component is heavily saturated in the 1-halo regime, and correcting for this saturation could introduce poorly constrained systematics into the measurement of $\xsgc$. A further complication arises from the overlap between the two doublet components. Fortunately, Figure~\ref{fig:stack} shows that the absorption becomes negligible beyond $|\Delta \vlos| \gtrsim 800 \, \mathrm{km \, s^{-1}}$ (corresponding to $|\Delta \lambda| \gtrsim 7.5 \, \textup{\AA}$) from the center of each component. We can therefore treat the outer half of each component as an isolated absorption profile, avoiding the need to deblend the overlapping inner halves. We consequently adopt the redward half of the weaker $\lambda2803$ component for our fiducial stacked measurement of $\xsgc$. Appendix~\ref{app:doublet} provides a more detailed justification for this choice.

Figure~\ref{fig:f2xi} demonstrates the efficacy of our conversion method, using the $0.01 < r_p \leq 0.5 \, h^{-1} \mathrm{Mpc}$ bin as an example. In the left panel, dashed gray histogram with uncertainty bands shows the composite absorption profile of the doublet, with the thicker black segments indicating the wavelength window used to derive the stacked $\xi^s(r_\pi \mid r_p)$ profile. In the right panel, we apply Equation~\eqref{eqn:xi^s} to convert the redward stacked $\lambda2803$ absorption profile into the redshift-space LRG-cloud cross-correlation function, $\xi^s(r_\pi \mid r_p)$, shown as open black circles with errorbars. For comparison, the open gray squares with errorbars show the $\xi^s(r_\pi \mid r_p)$ profile measured from the discrete \ion{Mg}{2} absorbers in the ancillary catalog. The stacked $\xsgc$ and discrete $\xsga$ profiles shares the same overall features, both including a narrower primary component and a broader secondary one.

Interestingly, the secondary component is more prominent in our fiducial $\xi^s$ profile than in the discrete one, producing a velocity hump between $400$ and $600 \, \mathrm{km \, s^{-1}}$. Since the primary component is the well-known FoG, this velocity hump resembles a ``knuckle'' superimposed on the FoG in the small-scale RSD signal. We have performed an extensive suite of tests and verified that this ``knuckle'' feature is not a statistical fluke (due to e.g., a few outliers), but a robust observational signature in the stacked profile (see Appendix~\ref{app:robust}).

\begin{figure*}
    \includegraphics[width=\linewidth]{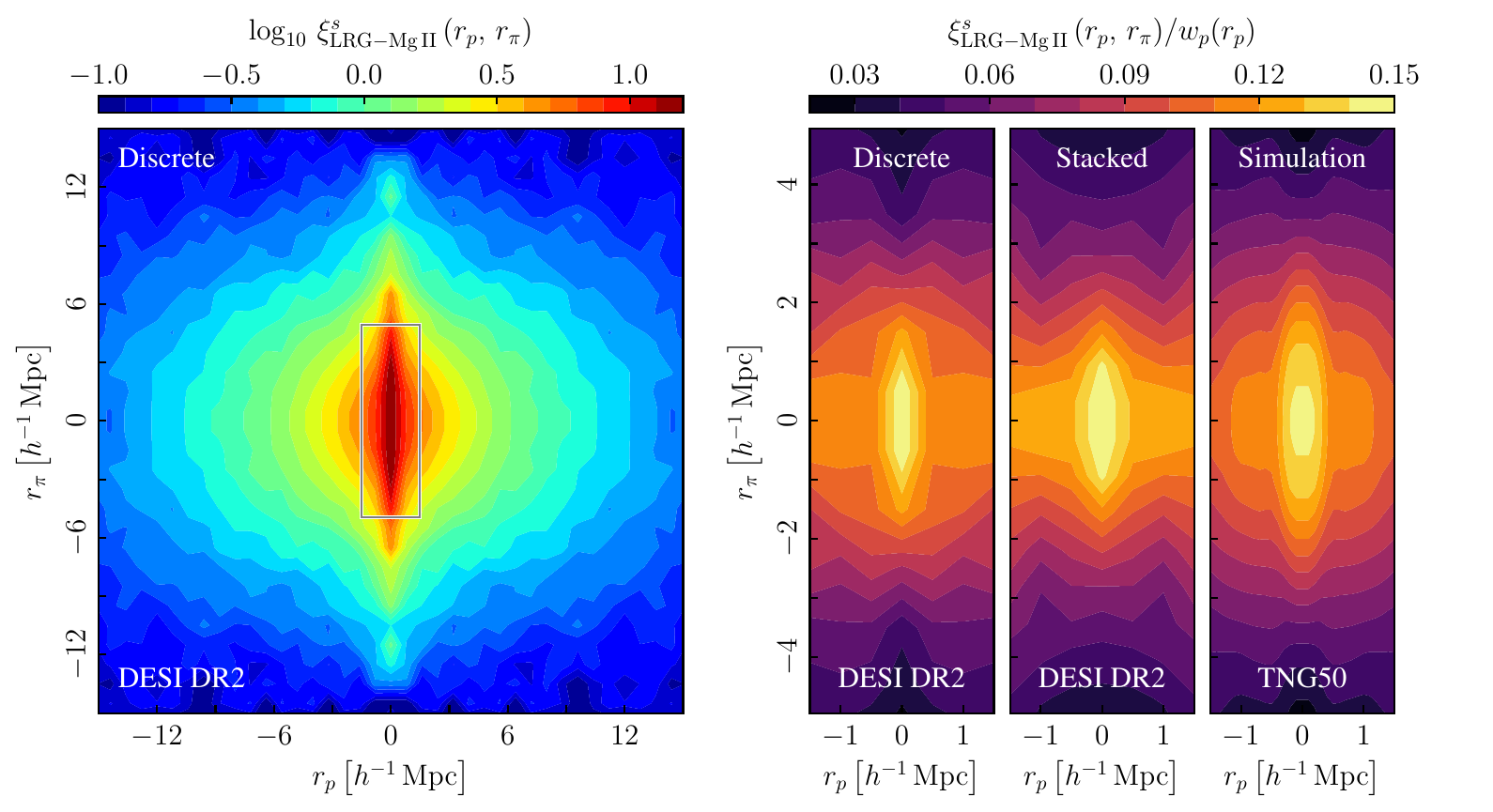}
    \caption{\textit{2D redshift-space LRG-\ion{Mg}{2} cross-correlation functions $\xi^s$.} \textbf{Left:} The $\xi^s(r_p, \, r_\pi)$ measurement using DESI LRGs and discrete \ion{Mg}{2} absorbers, displayed with a log-scale color map shown at the top. The gray rectangular box at the center has a dimension of $|r_p| < 1.5 \, h^{-1} \mathrm{Mpc}$ and $|r_\pi| < 5 \, h^{-1} \mathrm{Mpc}$, highlighting the FoG effect within the halo. \textbf{Right panels:} Zoomed-in views of the RSD pattern within the gray bounding box in the left panel. Filled contours indicate $\xi^s(r_p, \, r_\pi) / w_p(r_p)$, color-coded by the colorbar at the top. The three columns show results derived from DESI discrete absorbers (left), DESI stacked spectra (middle), and \ion{Mg}{2} gas in the TNG50 simulation (right).}
    \label{fig:rsd}
\end{figure*}

Figure~\ref{fig:rsd} presents an overview of the 2D redshift-space LRG-\ion{Mg}{2} cross-correlation function measured from DESI DR2. To set the stage, we show the $\xsga(r_p, \, r_\pi)$ measured from discrete absorbers in the left panel, color-coded by the colorbar on the top. Compared to the SDSS-based $\xsga$ measurement from \citetalias{Zu2021} (their Fig.\ 3), the DESI DR2 presents a much clearer RSD signature --- the FoG effect appears as an elongation along the LOS at the innermost $r_p$, while the Kaiser effect~\citep{Kaiser1987} is evident at large $r_p$. Similar to the SDSS-based finding in \citetalias{Zu2021}, we see an abrupt FoG cutoff near $r_\pi \sim 13 \, h^{-1} \mathrm{Mpc}$ (i.e., $\Delta \vlos \sim 1{,}200 \, \mathrm{km \, s^{-1}}$).

In the right panel of Figure~\ref{fig:rsd}, we focus on the FoG feature in the 1-halo regime by zooming in on the gray rectangle region $|r_p| < 1.5 \, h^{-1} \mathrm{Mpc}$, ($|r_\pi| < 5 \, h^{-1} \mathrm{Mpc}$) of the left panel. The three sub-panels present the contours of $\xi^s(r_p, \, r_\pi) / w_p(r_p)$ for the discrete (left), stacked (middle), and TNG50 (right) measurements, respectively. We plot $\xi^s / w_p$ instead of $\xi^s$ to highlight the kinematic feature by removing the strong trend of $\xi^s$ with $r_p$~\citep{ZW2013}. For the TNG50 calculation, we use all \ion{Mg}{2}-bearing gas cells around LRG-analogs (described later below) in the $z = 0.5$ snapshot.

All three measurements share a similar oval structure in their central RSD patterns, but with varying degrees of elongation driven by different kinematics of the underlying tracer population. Beyond the central region along the LOS, the RSD signals that correspond to the high-velocity gas clouds decrease rapidly at $|r_\pi| > 3 \, h^{-1} \mathrm{Mpc}$ for the two real observations. Thanks to the spectral stacking scheme, however, our fiducial $\xsgc$ measurement (middle) has a signal-to-noise ratio that is higher by a factor of $\sim 1.7$ than the discrete version (left), which would help reveal any secondary kinematic features like the ``knuckle'' at large $r_\pi$. Meanwhile, the TNG50 measurement has a much more extended FoG along the LOS than the two real observations, suggesting that cool cloud kinematics around the simulated LRGs is ``hotter'' than around the DESI ones.

Despite the stronger elongation, the RSD pattern in the TNG50 simulation resembles a rescaled version of that in the DESI stacked measurement. This resemblance is very encouraging, suggesting that the TNG50 likely captures the key features of cool cloud kinematics in the DESI data. In the next Section, we look to the TNG50 simulation for insights on how to construct a realistic kinematic model for the DESI \ion{Mg}{2} absorbing clouds.

\section{A TNG50-Guided Model of Cool Cloud Kinematics}
\label{sec:tng}

Cosmological hydrodynamical simulations explicitly track the baryon cycle through the CGM, thereby providing a realistic template for modeling cool cloud kinematics. Below we briefly describe the construction of mock LRG and \ion{Mg}{2}-absorbing cloud samples from the TNG50-1 simulation. Impatient readers can skip this Section directly to the theoretical model presented in \S~\ref{subsec:tng50_vdf}.

\subsection{Mock Samples of LRG and \texorpdfstring{\ion{Mg}{2}}{Mg II} Ions}
\label{subsec:mock}

The DESI LRGs are massive quiescent galaxies with a comoving number density of $\bar n_g \simeq 6 \times 10^{-4} \, h^{3} \mathrm{Mpc}^{-3}$ at $z \simeq 0.58$~\citep{Zhou2020}. To mimic such a sample, we select central galaxies with specific star formation rate $\mathrm{sSFR} < 10^{-10} \, \mathrm{yr}^{-1}$ from the $z = 0.5$ snapshot as our mock LRGs and threshold the mock sample by stellar mass to the observed $\bar n_g$. This selection of mock LRGs yields an average log-stellar mass of $11.50$, in good agreement with the DESI LRGs~\citep[11.31][]{Chen2025}; while the average log-halo mass (12.91; with a virial velocity dispersion of $\sigma_\mathrm{vir} = 222 \, \mathrm{km \, s^{-1}}$) is substantially lower than the observational constraints~\citep[$13.40 \pm 0.02$][]{Yuan2024}.

To generate mock \ion{Mg}{2} clouds, we post-process the simulation output at $z = 0.5$ to predict \ion{Mg}{2} fractions for individual gas cells using \textsc{Trident}~\citep{Hummels2017}. \textsc{Trident} tabulates ion fractions from single-zone \textsc{Cloudy} models~\citep{Ferland2013} under the~\citet{HM2012} UV background, over a 4D grid of hydrogen density $n_\mathrm{H}$, gas temperature $T$, gas metallicity $Z$, and redshift $z$. We interpolate this table to obtain the \ion{Mg}{2} fractions for all TNG50 gas cells, thereby populating the entire snapshot with \ion{Mg}{2} ions.

\subsection{Dynamical Classification of the CGM Gas in TNG50}
\label{subsec:class}

\begin{figure}
    \includegraphics[width=\columnwidth]{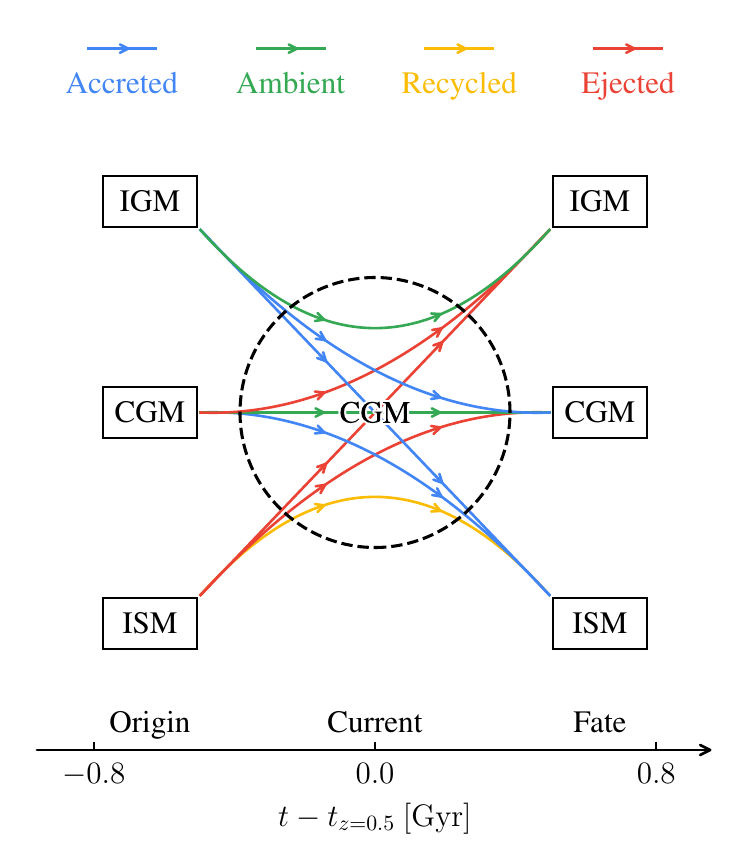}
    \caption{\textit{Dynamical classification of CGM gas cells in the TNG50 simulation.} The cartoon illustrates our method for classifying CGM gas cells at $z = 0.5$ (``Current'') into four dynamical categories (\emph{Accreted}, \emph{Ambient}, \emph{Recycled}, and \emph{Ejected}) based on their ``Origin'' and ``Fate'' states. The two end states are defined as the last and next non-CGM states before and after the time window centered at $z = 0.5$ ($1.6 \, \mathrm{Gyr}$), respectively.}
    \label{fig:cartoon}
\end{figure}

For the \ion{Mg}{2}-bearing gas cells in the LRG CGM, we identify their dynamical origin by tracking their evolution across different snapshots of the simulation. A similar strategy has been adopted by previous studies, e.g., \citet{Decataldo2024}. Although gas cells in \textsc{Arepo} undergo continuous refinement and de-refinement operations, their evolution can still be tracked with the Lagrangian tracer particles~\citep{Genel2013, Nelson2013}. These tracers track the volumetric flow, allowing reconstruction of histories for individual gas cells. For computational efficiency, we randomly select a subsample of $115{,}808$ \ion{Mg}{2}-bearing gas cells (consisting of $139{,}532$ tracers) from the CGM of our mock LRGs. We track the individual tracers of this subsample, recording their positions, velocities, parent particle types, and hydrogen densities across 60 simulation snapshots between $z = 1.5$ and $z = 0$.

At each snapshot, we classify gas cells into three types of ``state'' based on their galactocentric distance relative to the virial radius ($\tilde r \equiv r / \rvir$) and hydrogen density ($n_\mathrm{H}$):
\begin{itemize}[leftmargin=17pt, itemsep=2.5pt, parsep=2.5pt]
    \item Interstellar medium (ISM) state: $\tilde r \leq 0.1$;
    \item CGM state: $0.1 < \tilde r \leq 1$ and $n_\mathrm{H} \leq 0.106 \, \mathrm{cm}^{-3}$;
    \item IGM state: $\tilde r > 1$ and $n_\mathrm{H} \leq 0.106 \, \mathrm{cm}^{-3}$.
\end{itemize}
The $n_\mathrm{H}$ density threshold of $0.106 \, \mathrm{cm}^{-3}$ is set according to the two-phase star formation model of~\citealt{SH2003}. Within the tracking interval, a small fraction of tracers temporarily reside in non-gaseous resolution elements (e.g., star particles, stellar-feedback wind particles, or black hole sink particles). We group these tracers into the ISM because they are either spawned from or ejected out of star-forming gas cells. There also exists a tiny fraction ($< 1 \%$) of tracers that is associated with the satellite ISM ($\tilde r > 0.1$ and $n_\mathrm{H} > 0.106 \, \mathrm{cm}^{-3}$) and is therefore excluded from our analysis.

Gas cells frequently change their states due to baryon cycling. To identify the dynamical origin of each gas cell ``currently'' ($z = 0.5$) in the CGM state, we introduce a time window of $\simeq 1.6 \, \mathrm{Gyr}$ between $z_\mathrm{start} = 0.62$ and $z_\mathrm{end} = 0.40$, so that the immediate non-CGM snapshot before $z_\mathrm{start}$ and after $z_\mathrm{end}$ marks the tracer's ``origin'' and ``fate'' states, respectively. If no non-CGM stage is encountered, the origin or fate is set to the CGM state. Depending on the origin and fate states, we group all the current CGM gas cells into four dynamical classes, including \emph{Accreted}, \emph{Ambient}, \emph{Recycled}, and \emph{Ejected}, as indicated by the blue, green, yellow, and red curved arrows in the cartoon diagram of Figure~\ref{fig:cartoon}.

\subsection{Insight on the Cool Cloud Kinematics from TNG50}
\label{subsec:tng50_class}

\begin{figure*}
    \includegraphics[width=\linewidth]{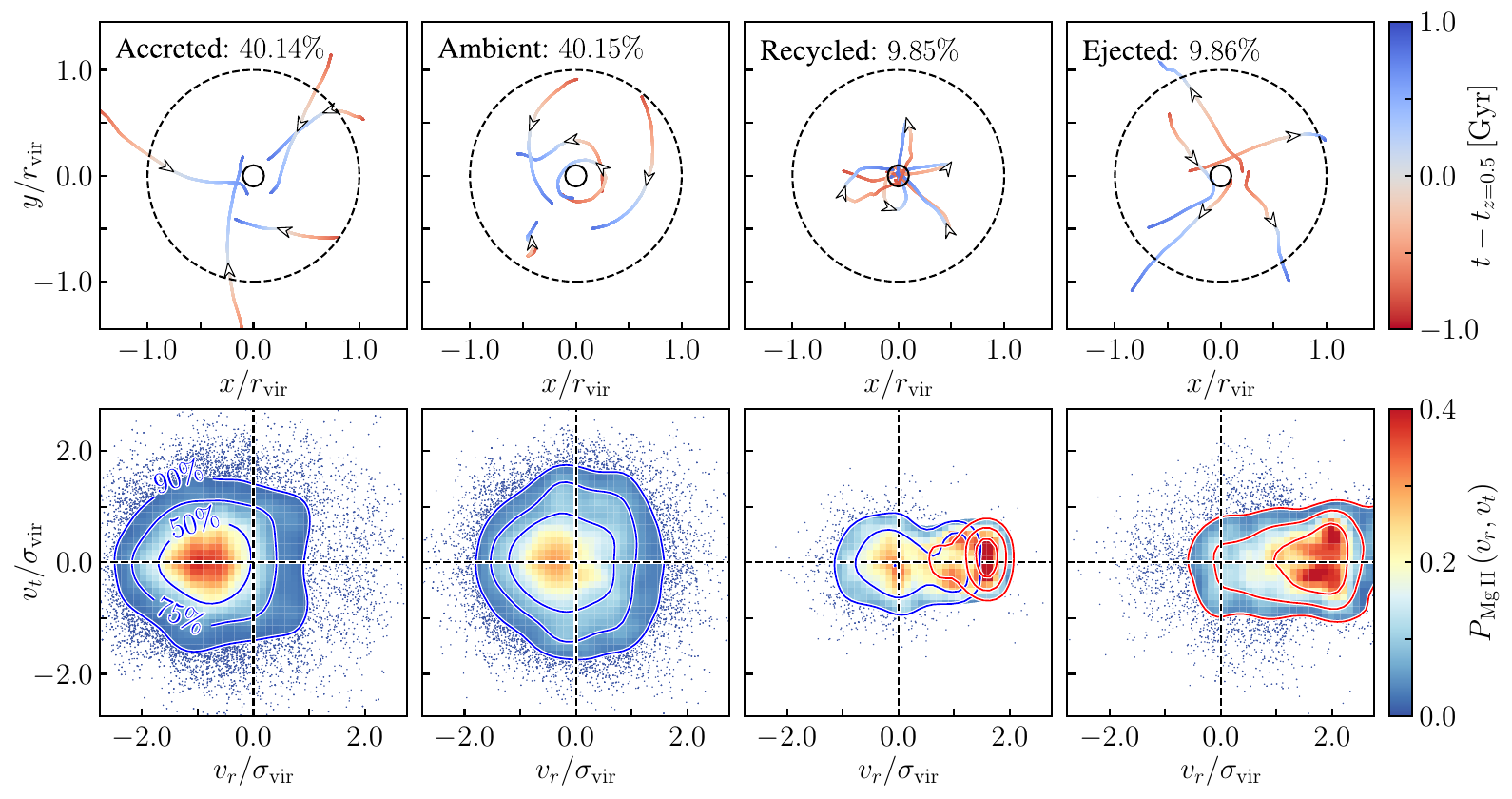}
    \caption{\textit{Cool gas kinematics in the TNG50 simulation.} \textbf{Top panels:} Solid colored curves with arrows are the five representative gas trajectories in the rest-frame of the LRGs, randomly selected from each of the four dynamical classes (with the percentage of that class denoted in the upper left of each panel). The arrow on each curve indicates the particle position and velocity direction at $z = 0.5$, while the color gradient along the curve indicates the time epoch relative to $z = 0.5$, according to the colorbar on the far right. Large dashed and small solid circles represent the average virial radius and the ISM boundary ($\tilde r = 0.1$) surrounding the LRGs, respectively. \textbf{Bottom panels:} Filled contours indicate the 2D VDF of \ion{Mg}{2} gas from each dynamical class, $P_\mathrm{Mg\,\textsc{ii}}(v_r, \, v_t)$, color-coded according to the colorbar on the far right, while the contour lines mark the regions that enclose $50$, $75$, and $90\%$ of the gas cells. Red and blue contour lines in the third panel (\emph{Recycled}) indicate the separate VDFs of the ``Fast-Recycled'' and ``Slow-Recycled'' subcomponents, respectively (see text). The radial (tangential) velocities on the $x$($y$)-axis are normalized by the virial velocities of halos.}
    \label{fig:tracking}
\end{figure*}

The top panels of Figure~\ref{fig:tracking} showcase trajectories for five typical gas cells in each of the four dynamical classes. Each trajectory starts from the ``origin'' at $z_\mathrm{start}$ (red), reaches the ``current'' position (white) with its velocity direction indicated by the open black arrow, and ends at the ``fate'' position at $z_\mathrm{end}$ (blue). We use a 6D phase-space interpolation~\citep{Patton2024} across different snapshots to smooth the paths for visualization, with the curves color-coded by the time since $z = 0.5$ according to the colorbar on the far right. Black dashed circles indicate $\tilde r = 1$. Overall, our simple scheme of categorizing CGM gas cells works surprisingly well, as the \emph{Accreted}, \emph{Ambient}, \emph{Recycled}, and \emph{Ejected} trajectories correctly reflect their respective dynamical properties within the $2 \, \mathrm{Gyr}$ span. At $z = 0.5$, the CGM mass budget is dominated by the \emph{Accreted} and \emph{Ambient} classes, each of which constitutes roughly $40\%$ of the gas cells, with the remaining $20\%$ shared equally by the \emph{Recycled} and \emph{Ejected} classes.

The bottom panels of Figure~\ref{fig:tracking} present the 2D\footnote{Following~\citealt{ZW2013}, we define the tangential velocity as $v_t = |\mathbf{v}_T|\cos\psi$, where $\mathbf{v}_T=\mathbf{v}-v_r\,\hat{\mathbf r}$ is the transverse component in the plane spanned by a random line of sight and the separation vector, $\psi$ is drawn uniformly in $[-180^\circ, 180^\circ]$ assuming isotropy.} velocity distribution functions (VDFs) $p(v_r, \, v_t)$ (weighted by the \ion{Mg}{2} mass) of CGM gas from the four corresponding dynamical classes shown in the top. In each bottom panel, the 2D VDF values are color-coded according to the colorbar on the right. In the leftmost panel, the radial velocities of the \emph{Accreted} CGM gas cells are mostly negative, indicating strong infall with a typical speed about half of the virial velocity. The second panel (from the left) shows that the 2D VDF of the \emph{Ambient} CGM gas cells is largely centered around $v_r = 0$, but the marginal distribution of $v_r$ is slightly skewed toward negative velocities. This skewness likely arises from the down-spiraling motion of \emph{Ambient} CGM gas cells as they gradually lose angular momentum within the halo. In the third panel, the 2D VDF of the \emph{Recycled} CGM gas has two velocity components, with one component centered on $v_r = 0$ (blue contours) and the other on $v_r \sim 1.2 \, \sigma_\mathrm{vir}$ (red contours). The latter component primarily consists of outflowing gas cells that were driven by galactic winds launched in the latest star-forming episode. Finally, in the rightmost panel, the \emph{Ejected} gas cells have a single outflowing component (red contours) that peaks around $1.5 \, \sigma_\mathrm{vir}$ but with a long tail reaching into $v_r \sim 0$. This component likely originates from the same winds that drove the outflowing component of the \emph{Recycled} gas. The overall consistency between the individual trajectories (top row) and the 2D VDFs (bottom row) exhibited by Figure~\ref{fig:tracking} reinforces the confidence in our simple scheme of dynamical classification.

After carefully examining the top panels of Figure~\ref{fig:tracking}, we arrive at a key insight for the modeling of cool cloud kinematics: despite their complex dynamical origins, the four dynamical classes of CGM gas can be effectively grouped into two kinematic components, one inflowing and the other outflowing. Specifically, we group \emph{Accreted} and \emph{Ambient} into the \emph{Inflow} component, and \emph{Ejected} gas into the \emph{Outflow} one. For the \emph{Recycled} class, we divide it into the ``Fast-Recycled'' (time spent in the CGM $T_\mathrm{CGM} < 0.45 \, \mathrm{Gyr}$) and ``Slow-Recycled'' ($T_\mathrm{CGM} \geq 0.45 \, \mathrm{Gyr}$) subcomponents. The 2D VDFs of the ``Fast-Recycled'' and ``Slow-Recycled'' components are indicated by the red and blue contour lines, respectively, in the bottom third panel of Figure~\ref{fig:tracking}. Consequently, the ``Slow-Recycled'' ($3.18\%$) CGM gas, which shows more virialized random motions, is grouped with \emph{Inflow}, while the ``Fast-Recycled'' ($6.67\%$) CGM gas is grouped with \emph{Outflow}.

\subsection{2D Velocity Distribution of CGM Gas in TNG50}
\label{subsec:tng50_vdf}

\begin{figure}
    \includegraphics[width=\columnwidth]{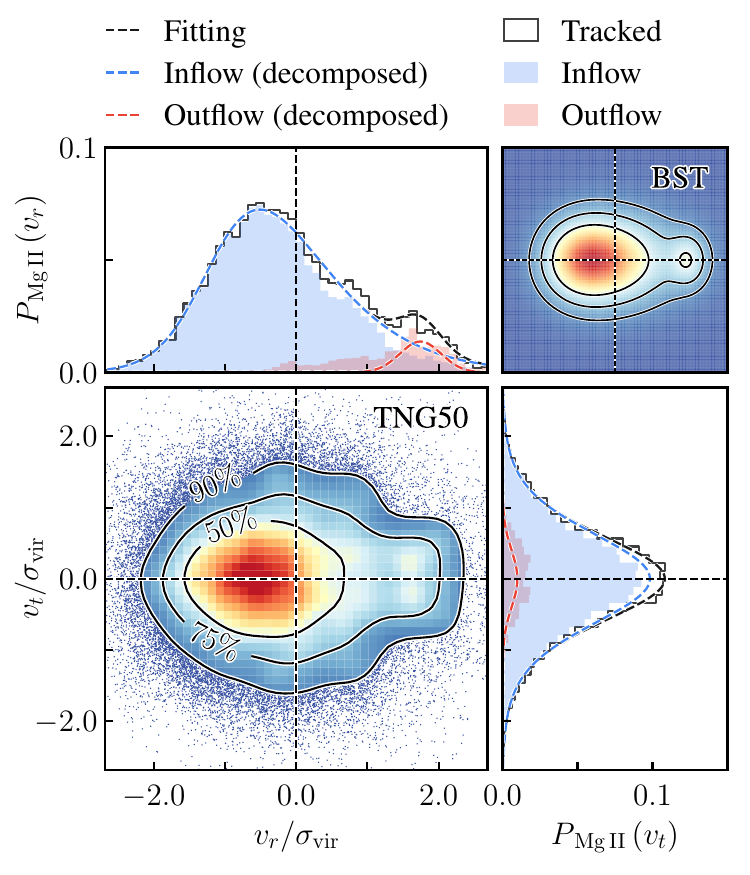}
    \caption{\textit{The bimodal skewed-t model for the 2D VDF of cool CGM gas surrounding LRGs in the TNG50 simulation.} The 2D VDF for all tracked gas particles (weighted by \ion{Mg}{2} mass) is shown in the lower-left panel, while the best-fitting BST model is displayed in the upper-right panel. Marginalized 1D VDFs of $v_r$ and $v_t$ are shown in the upper-left and lower-right panels, respectively. In these two panels, solid black histograms are directly measured from the simulation, and dashed black, blue, and red curves are predicted by the best-fitting BST model for the total, inflow and outflow components, respectively. For comparison, blue (red) filled histograms represent the VDFs directly measured from the \emph{Inflow} (\emph{Outflow}) gas cells classified by our particle tracking method.}
    \label{fig:vdf_2d}
\end{figure}

We now seek to develop an analytic model of cool cloud kinematics for describing the \emph{Inflow} and \emph{Outflow} components identified in TNG50. The main panel (lower left) of Figure~\ref{fig:vdf_2d} shows the 2D VDF of all CGM gas cells, which exhibits a bimodal structure including a pronounced, dispersed inflow and a less prominent but more collimated outflow. To model this bimodality, we follow~\citet{ZW2013} and adopt a two-component mixture of independent 2D skewed-$t$ distributions~\citep{AC2009}. This 2D bimodal Student-$t$ (hereafter referred to as ``BST'') model describes the VDFs of the \emph{Inflow} and \emph{Outflow} components using two separate 2D skewed-$t$ PDFs. Defining
$\mathbf{v} = (v_r, \, v_t)^\mathsf{T}$ and $\bar{\mathbf{v}} = (v_{r, \, \mathrm c}, \, 0)^\mathsf{T}$, we write the 2D skewed-$t$ PDF as
\begin{equation}
    \mathcal{T}(\mathbf{v}) = \, 2 \, t_2(\mathbf{v} - \mathbf{\Bar{v}}; \, \nu, \Sigma) \times T\left[ \tilde{v}_r; \, \nu + 2 \right] ,
\end{equation}
where $t_2$ is the bivariate Student's-$t$ distribution with degrees of freedom $\nu$ and covariance $\Sigma = \mathrm{diag}(\sigma_r^2, \, \sigma_t^2)$, $T$ is the cumulative distribution function of univariate Student's-$t$, and
\begin{equation}
    \tilde{v}_r = \frac{\alpha_r \, (v_r - v_{r, \, \mathrm c})}{\sigma_r} \left( \frac{\nu + 2}{Q_v + \nu} \right)^{1 / 2} ,
\end{equation}
with $Q_v = (\mathbf{v} - \mathbf{\Bar{v}})^\mathsf{T} \, \Sigma^{-1} \, (\mathbf{v} - \mathbf{\Bar{v}})$. The full 2D BST model is therefore
\begin{equation}
    P(\mathbf v) = f_\mathrm{in} \, \mathcal T_\mathrm{in}(\mathbf v) + f_\mathrm{out} \, \mathcal T_\mathrm{out}(\mathbf v) ,
    \label{eqn:bst}
\end{equation}
where $f_\mathrm{in}$ and $f_\mathrm{out} = 1 - f_\mathrm{in}$ are the fractional contributions of the two components, while $\mathcal T_\mathrm{in}(\mathbf v)$ and $\mathcal T_\mathrm{out}(\mathbf v)$ are the two corresponding 2D skewed-$t$ distributions.

We fit the measured 2D VDF in the main panel of Figure~\ref{fig:vdf_2d} with the 2D BST model of Equation~\eqref{eqn:bst}, using the Poisson likelihood
\begin{equation}
    -\ln\mathcal L = \sum_{i, j} \left[ \lambda_{ij} - n_{ij} \ln \lambda_{ij} + \ln \Gamma(n_{ij} + 1) \right] ,
\end{equation}
where $i$ and $j$ are the bin indices in $v_r$ and $v_t$, $\lambda_{ij}$ is the expected number of gas cells in the 2D bin $ij$ predicted by Equation~\eqref{eqn:bst}, and $n_{ij}$ is the number observed in the simulation. Note that both velocities are normalized by the halo virial dispersion $\sigma_\mathrm{vir}$ ($\langle \sigma_\mathrm{vir} \rangle = 202 \, \mathrm{km \, s^{-1}}$). The best-fitting model (top right panel of Figure~\ref{fig:vdf_2d}) agrees reasonably well with the 2D VDF measured from the simulation, with the maximum likelihood parameters being $\{v_{r, \, \mathrm c}, \, \sigma_r, \, \sigma_t, \, \alpha_r, \, \nu\} = \{-238.4, 269.7, 134.5, 2.0, 10.0\}$ for $\mathcal T_\mathrm{in}$, $\{343.4, 59.6, 81.6, 0.2, 201.4\}$ for $\mathcal T_\mathrm{out}$, and $f_\mathrm{in} = 0.96$.

In the top (right) side panel, we compare the marginalized 1D velocity distributions of $v_r$ ($v_t$) between the simulation measurement and model prediction, shown by the black histogram and black dashed curves, respectively. As discovered in \S~\ref{subsec:tng50_class}, the simulated CGM kinematics comprise \emph{Inflow} (\emph{Accreted}, \emph{Ambient}, and ``Slow-Recycled'') and \emph{Outflow} (``Fast-Recycled'' and \emph{Ejected}) components, shown as blue and red histograms, respectively. The best-fitting BST model is similarly decomposed into its \emph{Inflow} (blue dashed curve) and \emph{Outflow} (red dashed curve) contributions. We emphasize that these dashed curves are not direct fits to the histograms; rather, the consistency between each colored curve and its corresponding histogram demonstrates the success of our 2D BST model in describing the complex kinematics of cool clouds predicted by the TNG50.

\section{Observational Constraint on the Cool Cloud Kinematics}
\label{sec:fog}

\begin{figure*}[ht!]
    \includegraphics[width=\linewidth]{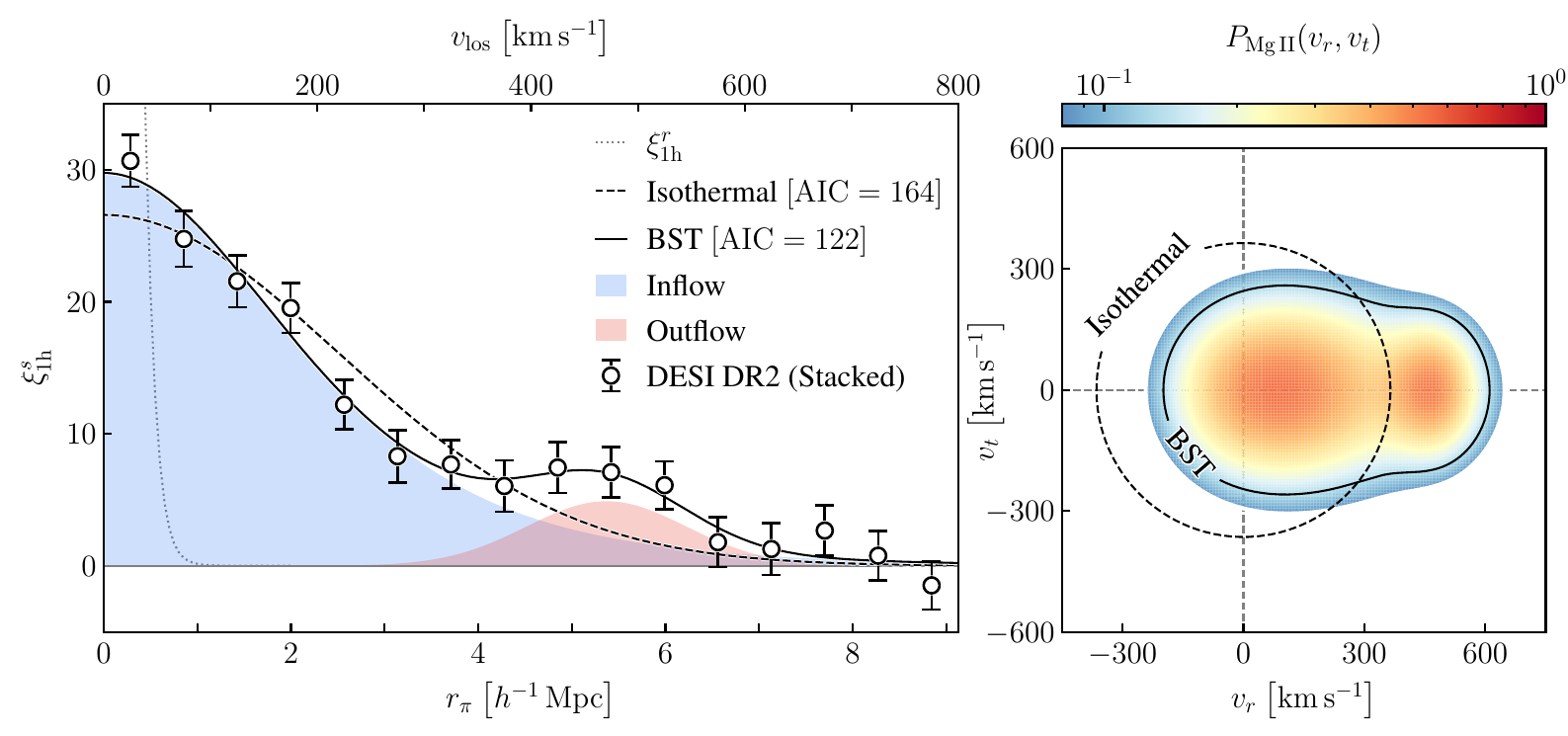}
    \caption{\textit{The ``knuckle'' in the FoG as a possible signature of outflows in the observed RSD of \ion{Mg}{2} gas around DESI LRGs.} \textbf{Left:} Black open circles with errorbars indicate our fiducial stacked measurement of the redshift-space LRG-\ion{Mg}{2} cross-correlation function within the halo, $\xsOh(r_\pi)$. Gray dotted curve is the real-space LRG-\ion{Mg}{2} cross-correlation function deprojected from the observed REW profile, $\xrOh(r)$. Solid black curve is the prediction by our best-fitting BST model, which is decomposed into an \emph{Inflow} component (blue filled region) and an \emph{Outflow} (red filled region) one. Another prediction from a simple isothermal model is shown as the black dashed curve for comparison. \textbf{Right:} Both the filled color contours and solid black contour line indicate the best-fitting BST 2D VDF of our cool CGM kinematics, including an \emph{Inflow} component and an \emph{Outflow} one. Black dashed contour line indicates the prediction by the isothermal velocity model. Both contour lines enclose $75\%$ of the respective populations.}
    \label{fig:fog}
\end{figure*}

\subsection{Derivation of the \texorpdfstring{\ion{Mg}{2}}{Mg II} RSD within the LRG Halos}
\label{subsec:lrg_mg2}

Armed with the 2D BST model inspired by TNG50, we are now ready to interpret the complex \ion{Mg}{2} RSD signal $\xsgc(r_p, \, r_\pi)$ measured in \S~\ref{subsec:xi_s}, including both the intriguing ``knuckle'' and the extended FoG effect. We focus exclusively on the 1-halo term of the redshift-space correlation $\xsOh(r_\pi \mid r_p < \rvir)$, which directly manifests the velocity distribution of the cool CGM gas inside the LRG halos. In particular, we identify two adjacent $r_p$ bins inside ($r_{p, \, \mathrm{in}} \in [0.01, \, 0.5]$ with $\langle r_{p, \, \mathrm{in}} \rangle = 0.36 \, h^{-1} \, \mathrm{Mpc}$) and outside ($r_{p, \, \mathrm{out}} \in [0.5, \, 1.0]$ with $\langle r_{p, \, \mathrm{out}} \rangle = 0.84 \, h^{-1} \, \mathrm{Mpc}$) of $\rvir$, and write their $\xsgc(r_\pi)$ profiles as
\begin{equation}
    \xsgc(r_\pi \mid r_{p, \, \mathrm{in}}) = \xsOh(r_\pi \mid r_{p, \, \mathrm{in}}) + \xsTh(r_\pi \mid r_{p, \, \mathrm{in}}) ,
\end{equation}
and
\begin{equation}
    \xsgc(r_\pi \mid r_{p, \, \mathrm{out}}) = \xsTh(r_\pi \mid r_{p, \, \mathrm{out}}) ,
\end{equation}
respectively. Following~\citetalias{Zu2021}, we assume $\xsTh(r_\pi \mid r_{p, \, \mathrm{in}}) = \xsTh(r_\pi \mid r_{p, \, \mathrm{out}})$, and measure the 1-halo term of the redshift-space correlation as
\begin{equation}
    \xsOh(r_\pi \mid r_{p, \, \mathrm{in}}) = \xsgc(r_\pi \mid r_{p, \, \mathrm{in}}) - \xsgc(r_\pi \mid r_{p, \, \mathrm{out}}) .
    \label{eqn:xi_1h_redshift}
\end{equation}

In practice, we fit a ``powered-exponential'' curve
\begin{equation}
    \xi^s_\mathrm{pe}(r_\pi) = \xi^s_0 \, \exp \left\{ - \left| \frac{r_\pi}{r_{\pi 0}} \right|^\beta \right\} ,
    \label{eqn:pe}
\end{equation}
to the $\xsgc(r_\pi \mid r_{p, \, \mathrm{out}})$ measurement ($\{ \xi^s_0, \, r_{\pi 0}, \, \beta \} = \{15.19, \, 4.22, \, 0.88\}$) and subtract this best-fitting curve in Equation~\eqref{eqn:xi_1h_redshift}. The result is shown by the open black circles with errorbars in the left panel of Figure~\ref{fig:fog}. The most interesting feature in the observed $\xsOh(r_\pi)$ profile is the extended plateau between $r_\pi = 4 \, h^{-1} \mathrm{Mpc}$ and $6 \, h^{-1} \mathrm{Mpc}$, echoing the secondary feature seen in the stacked absorption profile (Figure~\ref{fig:f2xi}).

\subsection{Constraint on the Velocity Distribution of Cool CGM}
\label{subsec:fog_fit}

For constraining the cool cloud kinematics, we expand the convolution introduced in Equation~\eqref{eqn:r2s} as
\begin{equation}
    \xsOh(r_\pi) = \int^{+\infty}_{-\infty} \xrOh(y) \, f(\vlos \mid y) \, \mathrm d y ,
    \label{eqn:xsOh}
\end{equation}
where $\vlos = H(z) (r_\pi - y) / (1 + z)$, $\xrOh(r)$ is the 1-halo term of the 3D real-space correlation inferred using Equation~\eqref{eqn:xi_1h}, $\xsOh$ is the 1-halo term of the redshift-space correlation derived from Equation~\eqref{eqn:xi_1h_redshift}.

The LOS velocity distribution is predicted by the 2D BST model given the five VDF parameters $\{ v_{r, \, \mathrm c}, \, \sigma_r, \, \sigma_t, \, \alpha_r, \, \nu \}$ for both the inflow and outflow components. During the deconvolution, we hold fixed the six TNG50-measured values for the two skewness parameters $\alpha_r$, the two degrees of freedom $\nu$, and the two $v_r / v_t$ dispersion ratios, assuming the hydro simulation correctly reproduces the shape of the 2D VDF. This leaves five free parameters: the inflow fraction $f_\mathrm{in}$, the characteristic radial velocity $v_{r, \, \mathrm{in}}$ and dispersion $\sigma_{r, \, \mathrm{in}}$ of the inflow, and the characteristic radial velocity $v_{r, \, \mathrm{out}}$ and dispersion $\sigma_{r, \, \mathrm{out}}$ of the outflow.

We constrain the BST model parameters by comparing the prediction from Equation~\eqref{eqn:xsOh} to the stacked RSD measurement (open black circles with errorbars) in Figure~\ref{fig:fog}. In the left panel, the black solid curve indicates the prediction by the best-fitting BST model. Underneath this predicted curve, the \emph{Inflow} and \emph{Outflow} components are shown as light blue and red regions, corresponding to the extended FoG effect and the ``knuckle'', respectively. Our $1$-$\sigma$ constraints on the five BST model parameters are $f_\mathrm{in} = 0.73 \pm 0.04$, $v_{r, \, \mathrm{in}} = -68.3 \pm 28.1 \, \mathrm{km \, s^{-1}}$, $\sigma_{r, \, \mathrm{in}} = 318.8 \pm 16.5 \, \mathrm{km \, s^{-1}}$, $v_{r, \, \mathrm{out}} = 459.3 \pm 16.0 \, \mathrm{km \, s^{-1}}$, $\sigma_{r, \, \mathrm{out}} = 73.0 \pm 7.1 \, \mathrm{km \, s^{-1}}$, respectively.

For comparison, a simple isothermal fit with $\sigma_\mathrm{los} = 238 \, \mathrm{km \, s^{-1}}$ is overplotted as the black dashed curve. To compare the two best-fitting models quantitatively, we compute the Akaike Information Criterion~\citep[$\mathrm{AIC}$;][]{Akaike1974} defined as $\mathrm{AIC} = 2 k - 2 \ln \mathcal{L}_\mathrm{max}$, where $k$ is the number of free parameters and $\mathcal{L}_\mathrm{max}$ is the maximum likelihood value evaluated using the bootstrap covariance matrix. With five free parameters, the BST model provides a better fit to the observed $\xsOh$ than the isothermal model, lowering the chi-squared per degree of freedom from $\chi^2_\mathrm{dof} = 2.05$ to $0.55$. This improvement persists after accounting for the additional model complexity, with an Akaike Information Criterion difference of $\Delta\mathrm{AIC} = 42$ ($\mathrm{AIC} = 164$ and $122$ for the isothermal and BST models, respectively). In the BST framework, the ``knuckle'' in the stacked FoG can be interpreted by a population of outflowing CGM clouds traveling at $v_r \simeq 459 \, \mathrm{km \, s^{-1}}$ within the halo.

\subsection{Physical Implication: The Tired Wind Scenario}
\label{subsec:lit_imp}

\subsubsection{Are the \texorpdfstring{\ion{Mg}{2}}{Mg II} RSDs consistent between SDSS and DESI?}

In the SDSS-based study by~\citetalias{Zu2021}, the \ion{Mg}{2} RSD around SDSS LRGs also exhibits a velocity hump in the FoG (see their Figure 11). After examining four competing models, \citetalias{Zu2021} concluded that the \ion{Mg}{2} RSD is best matched by the combination of a random velocity component and an outflowing one. In particular, the \citetalias{Zu2021} result favors a decelerated wind-bubble traveling with $v_r \simeq 250 \, \mathrm{km \, s^{-1}}$ at radius $\sim 400 \, h^{-1} \mathrm{kpc}$ away from the SDSS LRGs.

Interestingly, our best-fitting BST model is consistent with this ``tired wind'' scenario proposed by~\citetalias{Zu2021}, except that our inferred wind bubble is significantly faster ($\sim 459 \, \mathrm{km \, s^{-1}}$) and closer to the galaxy ($\sim 263 \, h^{-1} \mathrm{kpc}$) --- a somewhat brisker wind. Here, we explore whether such a faster wind-bubble can be physically produced and maintained in the CGM of massive quiescent galaxies, and we attempt to reconcile the velocity difference between the two studies.

Observationally, fast outflows with wind speed around $1000 \, \mathrm{km \, s^{-1}}$ can be detected up to a few $100 \, \mathrm{Myr}$ following a starburst in recently quenched massive galaxies~\citep{Tremonti2007, Coil2011, Taylor2024}. The subsequent evolution of those outflows remains unknown, but the semi-analytic wind model proposed by~\citet{Lochhaas2018} predicts that the fossil wind-driven bubbles can linger in the CGM for $5$--$10 \, \mathrm{Gyr}$. This tired wind scenario successfully reproduces individual absorption systems around passive galaxies~\citep{Tumlinson2013, Keeney2017} and adequately describes the RSD signal observed by~\citetalias{Zu2021}. Putting all the pieces together, it is tempting to regard our faster outflows as the same wind bubbles as in \citetalias{Zu2021}, but observed at an earlier epoch.

Following~\citet{Lochhaas2018}, we expect the two cool wind-bubbles are at the momentum-driven phase, governed by
\begin{equation}
    \frac{\mathrm{d}}{\mathrm{d}t} (M_b v_b) = \dot{p}_w - \frac{G M_b M_{\mathrm{enc}}(r)}{r^2},
\end{equation}
where $M_b$ is the mass of the bubble, $v_b$ is the bubble velocity, $\dot{p}_w$ is momentum deposition rate of the wind driving the bubble, $M_{\mathrm{enc}}$ is the total enclosed mass within radius $r$ that is responsible for the gravitational deceleration. For the LRGs, we assume wind injection has already ceased ($\dot{p}_w = 0$). The velocity of the bubble therefore obeys
\begin{equation}
    \frac{\mathrm{d}v_b}{\mathrm{d}t} = -\frac{G M_{\mathrm{enc}}(r)}{r^2} - \frac{v_b}{M_b} \frac{\mathrm{d}M_b}{\mathrm{d}t},
\end{equation}
while gaining mass as it sweeps up the ambient CGM gas with density $\rho_{\mathrm{gas}}$
\begin{equation}
    \frac{\mathrm{d}M_b}{\mathrm{d}t} = 4\pi r^2 \rho_{\mathrm{gas}}(r) v_b.
    \label{eqn:dMt}
\end{equation}
Combining the two equations above and $v_b = \mathrm{d}r / \mathrm{d}t$, we have
\begin{equation}
    \frac{\mathrm{d}(v_b^2)}{\mathrm{d}r} = -\frac{2G M_{\mathrm{enc}}(r)}{r^2} - 2v_b^2 \frac{\mathrm{d}\ln M_b}{\mathrm{d}r},
\end{equation}
with the two boundary conditions set by $(r_1, \, v_1) = (263\,h^{-1}\,\mathrm{kpc}, \, 459 \, \mathrm{km \, s^{-1}})$ and $(r_2, \, v_2) = (400 \, h^{-1}\,\mathrm{kpc}, \, 250\,\mathrm{km \, s^{-1}})$. For the sake of simplicity, we approximate the shell mass growth as a power-law within an exponent $q$
\begin{equation}
    M_b(r) = M_1 \left( \frac{r}{r_1} \right)^q,
\end{equation}
where $M_1$ is the shell mass at $r_1$. After integration, the resulting wind velocity can be solved as
\begin{equation}
    v_b^2(r) = \dfrac{M_1^2 v_1^2 - 2 \displaystyle \int\nolimits_{r_1}^{r} M_b^2(r^\prime) \dfrac{G M_{\mathrm{enc}}(r^\prime)}{r^{\prime 2}} \,\mathrm{d}r^\prime}{M_b^2(r)}.
    \label{eqn:v_b}
\end{equation}
Assuming an NFW host dark matter halo profile (Table~\ref{tab:wp_fit}), we can solve for the shell mass growth from Equation~\eqref{eqn:v_b} using the two boundary conditions, yielding the best-fitting value of $q = 0.51$.

Therefore, starting from $v_1 = 459\,\mathrm{km \, s^{-1}}$ at $r_1$ (this work), the wind bubble needs to plow through the CGM material and grow its mass by roughly $24\%$, i.e.,
\begin{equation}
    \frac{M_b(r_2)}{M_b(r_1)} \simeq 1.24,
\end{equation}
in order to be decelerated to $v_2 = 250 \, \mathrm{km \, s^{-1}}$ at $r_2$~\citepalias{Zu2021}. The associated travel time between the two snapshots is
\begin{equation}
    \Delta t \simeq 0.59 \, \mathrm{Gyr}.
\end{equation}
This timescale is consistent with the period required for galaxies to undergo their final star formation episode and quench, potentially accompanied by episodic AGN activity as the starburst fades~\citep{Taylor2024}.

The mass-growth requirement also provides an interesting constraint on the ambient gas distribution. If the shell mass is dominated by swept-up gas, Equation~\eqref{eqn:dMt} implies that
\begin{equation}
    \rho_{\mathrm{gas}}(r) \propto r^{q - 3} \simeq r^{-2.49},
\end{equation}
roughly consistent with the NFW but steeper than a singular isothermal sphere. Alternatively, a shallower gas profile could reproduce the same deceleration if the shell already contains substantial accumulated mass at the smaller radius.

From the toy model calculation above, we show that the two observations are consistent with a tired wind-bubble slowing over roughly half a gigayear while growing one-fifth of its initial mass. Nevertheless, the inferred mass-loading depends on the assumed halo gas distribution, and a full reconstruction of the initial wind energetics would require modeling the earlier energy-driven phase and the subsequent cooling transition, which is beyond the scope of this paper.

\subsubsection{Are the wind bubbles bound to the LRG halos? }

Extending the momentum-conserving shell trajectory in Equation~\eqref{eqn:v_b} beyond $r_2$ under the same mass-sweeping law ($M_b \propto r^q$) and NFW potential, we find that gravitational deceleration from the host halo and the accumulated swept-up mass reduce the shell velocity to zero at $r_\mathrm{max} \simeq 0.97 \, \rvir$, after a propagation time of $\simeq 2.24 \, \mathrm{Gyr}$ from $r_1$. Therefore, the cool outflows would extend briefly beyond the virial boundary, stall, and eventually fall back while remaining gravitationally bound to the LRG halo.

Inferring the displacement of hot gas from LRGs is also vital to cosmology for understanding the impact of baryonic feedback on the matter power spectrum~\citep{Zhang2026}. In particular, we are primarily interested in the maximum radius of gas ejection $\theta_\mathrm{ej} \equiv r_\mathrm{max} / \rvir$~(normalized by $\rvir$)~\citep{Schneider2019} and the halo gas mass fraction $f_{\rm gas}$ within $\rvir$~(normalized by the cosmic baryon fraction). For example, \citet{Siegel2026} found a hot gas ejection parameter of $\theta_\mathrm{ej}\simeq 5{-}6$, using a joint analysis of kinetic Sunyaev--Zel'dovich~(kSZ) and X-ray observations.  This is significantly greater than our cool-shell turnaround at $\theta_\mathrm{ej} \simeq 0.97$, though joint constraints from kSZ and X-ray could be overestimated due to the degeneracy between baryonic feedback with satellite fraction~\citep{OM2026}. Using the stacked thermal SZ~(tSZ) observations, \citet{Chen2023} found that the gas fraction of LRG halos at $z\simeq 0.5$ is merely $\simeq 15\%$~(their Figure 8), which also requires a strong feedback that redistribute most of the gas outside $\rvir$~\citep{Bahar2024, Popesso2026, Khalil2026}. Therefore, the cool outflows traced by the \ion{Mg}{2} absorption are unlikely the culprit behind the observed high $\theta_\mathrm{ej}$ and low $f_{\rm gas}$. More energetic hot winds or powerful jets~\citep{Cen2024, Qiu2025} are required to effectively redistribute gas on scales beyond several $\rvir$, as seen in some hydrodynamic simulations~\citep{Wright2024}.

\subsubsection{The momentum budget of the outflows}

Although the shell trajectory in Equation~\eqref{eqn:v_b} depends only on relative mass growth, we can estimate the absolute radial momentum flux of the cool outflow by combining our measured \ion{Mg}{2} spatial profile $\xrOh(r)$ with the BST velocity distribution. To convert the \ion{Mg}{2} ion density into a cool-gas mass density $\rhcl$, we adopt $(\mathrm{Mg/H}) = 0.3 \, (\mathrm{Mg/H})_\odot$ with a \ion{Mg}{2} ionization fraction of $0.5$, a mean gas mass of $1.4 \, m_p$ per hydrogen nucleus, and negligible dust depletion. Evaluating the flow at $r_1 = 263 \, h^{-1} \mathrm{kpc}$ with the best-fitting outflow fraction $f_\mathrm{out} = 0.28$, we obtain
\begin{equation}
    \begin{aligned}
        \dot{p}_\mathrm{cool, \, out} & = 4\pi R_1^2 f_\mathrm{out} \rhcl(r_1) \langle u_r^2 \rangle_\mathrm{out} \\
        & \simeq 6.10 \times 10^{33} \, \mathrm{dyn} ,
    \end{aligned}
\end{equation}
where $\langle u_r^2 \rangle_\mathrm{out}$ is the second moment of the proper radial velocity distribution calculated from our BST model, and the inferred momentum flux scales inversely with the assumed magnesium abundance and ionization fraction. For a typical quasar-mode feedback, the characteristic momentum flux is
\begin{equation}
\dot p_{\rm rad}
=4.20\times10^{35}
    \left(\frac{\lambda_{\rm Edd}}{0.1}\right)
\left(\frac{M_\bullet}{10^9M_\odot}\right)
\ {\rm dyn},
\end{equation}
where $\lambda_{\rm Edd}$ is the Eddington ratio and $M_\bullet$ is the mass of the central supermassive black hole. More sophisticated AGN feedback schemes like the MACER~\citep{Yuan2018, Zhu2023, Zhang2025} would yield a similar momentum flux of $10^{35}\,\mathrm{dyn}$.

The inferred momentum flux of the cool outflow is substantially smaller than that injected by the AGN feedback. This difference is not necessarily inconsistent with the wind-bubble interpretation, as Mg II absorption traces only the cool component of a potentially multiphase outflow. Instead, the observed cool-gas momentum flux can be supplied by the last AGN episode, while the total momentum budget of the expanding bubble remains dependent on its hot-gas content and earlier energy-driven evolution, which can be potentially inferred from X-ray or tSZ observations.

\subsubsection{Dependence of outflow velocity on post-starburst age}

Finally, we design an experiment to test the tired wind scenario directly using the DESI observations. The DESI LRG sample is not a homogeneous population of quiescent galaxies, but contains a wide spread in stellar age since the last star-burst~\citep{Setton2023, Zhang2024}. We divide the DESI LRG sample into three subsamples based on their $4000 \, \textup{\AA}$ break indices ($D_n (4000)$), an observable quantity that is positively correlated with the post-starburst age~\citep{French2018}. In order to isolate the contribution of $D_n(4000)$, we further require the tertiles to have the same joint distribution of redshift and stellar mass. Afterward, we repeat the stacked RSD analysis for each of the three subsamples.

\begin{figure}
    \includegraphics[width=\columnwidth]{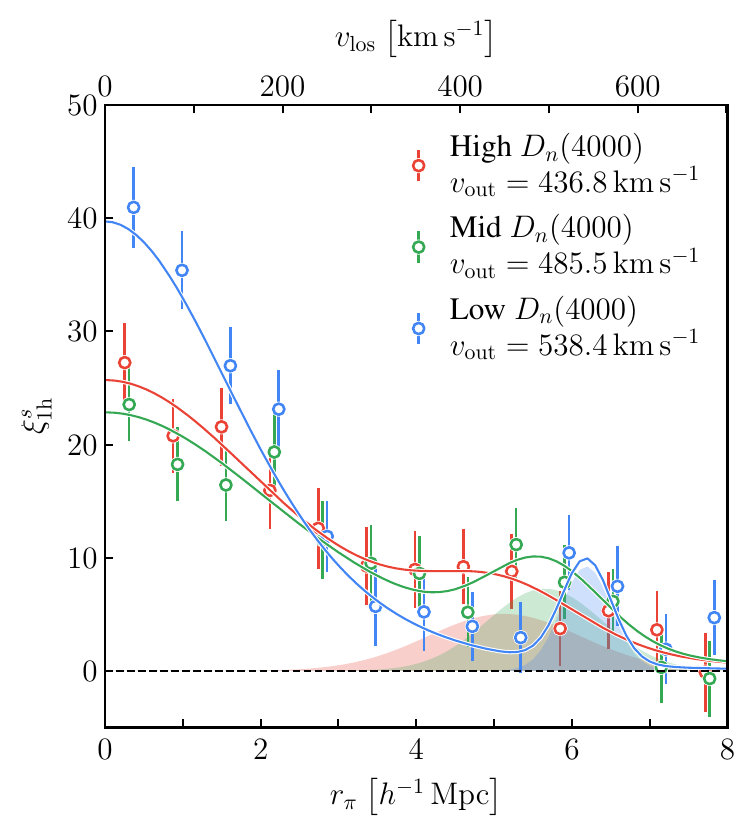}
    \caption{\textit{The stacked RSD measurements for three DESI LRG subsamples binned by their $D_n(4000)$.} Red, green, and blue circles with errorbars indicate the stacked measurements of $\xsOh$ for the high, mid, and low-$D_n(4000)$ subsamples, respectively. The data points at the same $r_\pi$ are slightly shifted horizontally for visual clarity. Solid curves in the corresponding colors represent the RSD prediction by the best-fitting BST models, while the red, green and blue filled regions highlight their respective outflowing components.}
    \label{fig:xi_dn4000}
\end{figure}

Figure~\ref{fig:xi_dn4000} presents the 1-halo RSD stacking results for these $D_n(4000)$-binned subsamples. By isolating the outflow contribution (color-filled regions), we highlight a clear monotonic decline in the velocity of the outflow component from the youngest (low $D_n(4000)$; blue) to the oldest (high $D_n(4000)$; red) subsample --- the ``knuckle'' rolls inwards with increasing $D_n(4000)$. Notably, the oldest subsample exhibits a much slower outflow, with an amplitude closer to the previous SDSS-based measurements. This behavior qualitatively aligns with~\citet{Sun2023}, who reported an exponential decline in outflow velocity as a function of post-starburst age. While further observations are required to confirm this interpretation, this $D_n(4000)$ dependence provides supporting evidence that the ``knuckle'' could be associated with the fossil wind left by the last episode of star-forming or AGN activities --- the last breath of massive quiescent galaxies.

\section{Summary}
\label{sec:summary}

In this paper, we present a novel technique to directly measure the 2D redshift-space cross-correlation $\xsgc$ between the LRGs and \ion{Mg}{2} absorbing cool gas using DESI DR2. By converting the stacked \ion{Mg}{2} absorption line profiles, $\fMg$, into $\xsgc$, we account for all the \ion{Mg}{2} absorption signals arising from the CGM, especially including those due to weak absorbers that are abundant but absent from discrete absorber catalogs. The RSD signal exhibits a complex RSD pattern that includes an intriguing ``knuckle'' in the usual FoG effect.

We formulate a bimodal skewed-$t$ (BST) model of cool CGM kinematics by examining the trajectories of \ion{Mg}{2}-bearing gas cells across cosmic time within the TNG50 hydro simulation. We demonstrate that this physically-motivated BST model provides an excellent description of our stacked measurement of $\xsgc$, revealing a dominant population ($72\%$) of inflowing gas with radial velocity dispersion of $382 \, \mathrm{km \, s^{-1}}$ and a secondary ($28\%$) outflowing component with a characteristic radial velocity of $459 \, \mathrm{km \, s^{-1}}$ and a small dispersion of $76 \, \mathrm{km \, s^{-1}}$. The two components explain the extended FoG effect and the ``knuckle'', respectively, in our stacked DESI RSD signal. Our main results are as follows:

\begin{enumerate}
    \item In the 1-halo regime ($r_p \lesssim 1\,h^{-1}\mathrm{Mpc}$), our stacked projected LRG-Mg II cross-correlation function $w_p(r_p)$ exceeds the traditional measurement using discrete absorbers by $\gtrsim 30 \%$. This discrepancy suggests that a substantial fraction of the cool CGM gas may reside in tiny fragments dispersed throughout the ambient medium, resembling the ``misty'' structure advocated by~\citet{McCourt2012}.

    \item The projected clustering $w_p(r_p)$ is well described by a two-component model with an NFW-like real-space 1-halo term $\xrgc$, consistent with a characteristic LRG mass of $\log_{10} \Mvir / \mathrm{M_\odot} = 13.07^{+0.16}_{-0.13}$.

    \item We observe complex RSD of \ion{Mg}{2} absorption around DESI LRGs in the stacked $\xsgc$ measurement. After the subtraction of the 2-halo contribution, the stacked 1-halo redshift-space cross-correlation, $\xsOh(r_\pi)$, exhibits a secondary velocity component at $\Delta \vlos \simeq 459 \, \mathrm{km \, s^{-1}}$ that is robust against different sample selections and stacking methods --- the ``knuckle'' in the FoG of CGM gas clouds.

    \item By convolving our best-fit $\xrgc$ with the TNG50-guided BST model, we successfully reproduce the observed RSD signal of \ion{Mg}{2} clouds around DESI LRGs, revealing an outflowing component traveling with a characteristic radial velocity of $459 \, \mathrm{km \, s^{-1}}$. This BST cool cloud kinematics model with outflows is favored by the DESI DR2 over a singular isothermal model, with a reduced chi-squared difference of $1.47$ and an Akaike Information Criterion difference of $40$, providing a plausible explanation for the ``knuckle''.

    \item After dividing the LRGs into three subsamples according to their $D_n(4000)$ indices, we find that the BST-inferred outflowing component persists in all subsamples. More interestingly, the inferred radial velocity decreases monotonically with the increase of $D_n(4000)$, a proxy for the time since the last star-forming episode of LRGs.
\end{enumerate}

These results demonstrate that spectral stacking serves as a highly sensitive probe of cool gas kinematics, capable of recovering weak absorptions typically overlooked by discrete absorber catalogs. By integrating the full absorption profile, stacking yields a comprehensive view of the \ion{Mg}{2}-bearing gas and its non-Gaussian, multi-component velocity structure.

However, although we have verified the robustness of this ``knuckle'' feature against various systematic tests (Appendix~\ref{app:robust}), its definitive physical origin remains an open question. Future data releases from DESI, along with complementary large-scale spectroscopic surveys such as the Prime Focus Spectrograph~\citep[PFS;][]{Takada2014, Greene2022}, the 4-meter Multi-Object Spectroscopic Telescope~\citep[4MOST;][]{deJong2019}, and Jiaotong University Spectroscopic Telescope~\citep[JUST;][]{JUST2024} will significantly increase the galaxy-quasar pair counts and improve constraints on the redshift-space galaxy-cloud cross-correlation signal. On the theoretical front, higher-resolution simulations capable of resolving the detailed phase structure of the cool CGM will be essential to validate the observed bimodal kinematics and link the high-velocity secondary component to specific feedback mechanisms.

\begin{acknowledgments}
    We thank Abhijeet Anand, Renyue Cen, Cameron Hummels, Rahul Ramesh, Ma\l{}gorzata Siudek, Xuanyi Wu, and Feng Yuan for helpful comments and discussions. This work is supported by the National Natural Science Foundation of China (12595313), the National Key Basic Research and Development Program of China (No. 2023YFA1607800, 2023YFA1607804), and the China Manned Space Program (No. CMS-CSST-2025-A04). Y.Z. acknowledges the generous sponsorship from Yangyang Development Fund. Y.Z. thanks Cathy Huang for her hospitality at the Zhangjiang High-tech Park.

    This material is based upon work supported by the U.S. Department of Energy (DOE), Office of Science, Office of High-Energy Physics, under Contract No. DE-AC02-05CH11231, and by the National Energy Research Scientific Computing Center, a DOE Office of Science User Facility under the same contract. Additional support for DESI was provided by the U.S. National Science Foundation (NSF), Division of Astronomical Sciences under Contract No. AST-0950945 to the NSF's National Optical-Infrared Astronomy Research Laboratory; the Science and Technology Facilities Council of the United Kingdom; the Gordon and Betty Moore Foundation; the Heising-Simons Foundation; the French Alternative Energies and Atomic Energy Commission (CEA); the Secretariat of Science, Humanities, Technology and Innovation (SECIHTI) of Mexico; the Ministry of Science, Innovation and Universities of Spain (MICIU/AEI/10.13039/501100011033), and by the DESI Member Institutions: \url{https://www.desi.lbl.gov/collaborating-institutions}. Any opinions, findings, and conclusions or recommendations expressed in this material are those of the author(s) and do not necessarily reflect the views of the U. S. National Science Foundation, the U. S. Department of Energy, or any of the listed funding agencies.

    The authors are honored to be permitted to conduct scientific research on I'oligam Du'ag (Kitt Peak), a mountain with particular significance to the Tohono O'odham Nation.
\end{acknowledgments}

\software{
    \textsc{Astropy}~\citep{Astropy2013,Astropy2018,Astropy2022}, \textsc{Cloudy}~\citep{Ferland2013}, \textsc{NumPy}~\citep{Numpy2020}, \textsc{SciPy}~\citep{SciPy2020}, \textsc{Trident}~\citep{Hummels2017}
}

\appendix
\section{Residual Continuum Correction in Spectral Stacking}
\label{app:stk_cor}

Composite absorption spectra are susceptible to systematic continuum residuals arising from continuum-fitting errors, data-reduction artifacts, and unmasked spectral features. To remove these residuals from our stacked \ion{Mg}{2} absorption profiles, we apply a two-step correction procedure.

\renewcommand\thefigure{A}
\begin{figure}
    \includegraphics[width=\columnwidth]{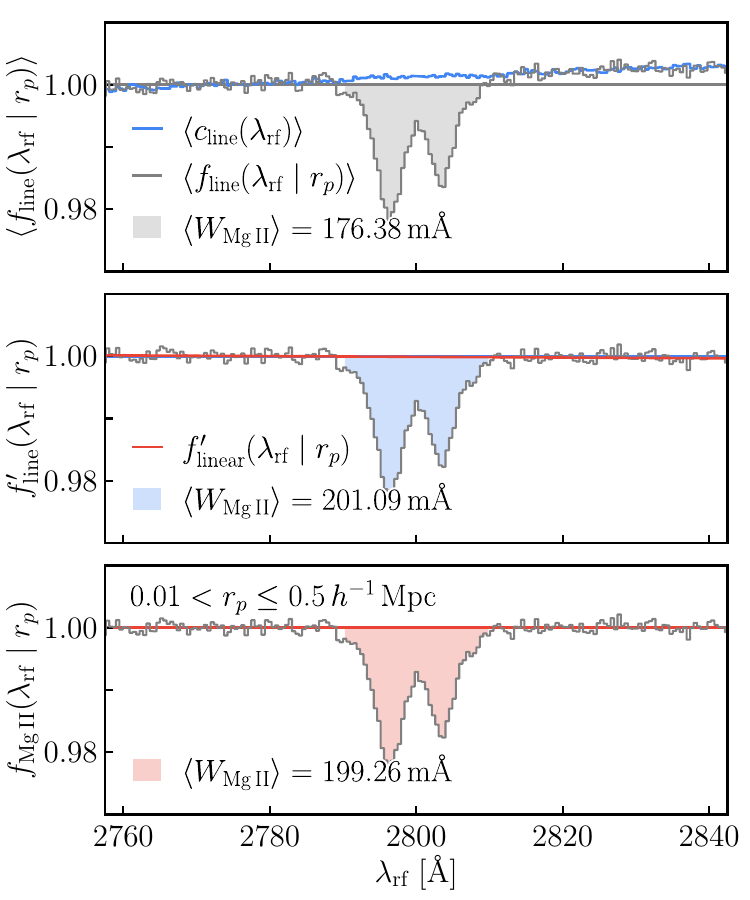}
    \caption{\textit{Two-step correction procedure for the stacked \ion{Mg}{2} absorption profile.} \textbf{Top:} The uncorrected stacked spectrum $\flin$ (gray solid histogram) alongside the random-quasar stacked continuum $\clin$ (blue solid histogram). \textbf{Middle:} The intermediate composite spectrum $f^\prime_\mathrm{line}$ (gray solid histogram) after normalization by $\clin$. The red solid line shows the best-fitting linear continuum $f^\prime_\mathrm{linear}$ obtained by masking out the wavelength range $2780$--$2820 \, \textup{\AA}$. \textbf{Bottom:} The final corrected \ion{Mg}{2} absorption profile $\fMg$ (gray solid histogram), derived by dividing $\flin$ by both $\clin$ and $f^\prime_\mathrm{linear}$. In all three panels, color-filled histograms indicate the integration windows and corresponding REW measurements.}
    \label{fig:stk_cor}
\end{figure}

Figure~\ref{fig:stk_cor} illustrates this two-step procedure using the $0.01 < r_p \leq 0.5 \, h^{-1} \mathrm{Mpc}$ bin as an example. The top panel compares the direct stacking result $\flin$ (gray solid histogram), constructed from the NMF-normalized quasar spectra $\fqso$, with the random-quasar stacked continuum $\clin$ (blue solid histogram). Outside the absorption window ($\lamr < 2780 \, \textup{\AA}$ and $\lamr > 2820 \, \textup{\AA}$), $\clin$ closely tracks the global offset of $\flin$ from unity. In the first step, we divide $\flin$ by $\clin$ to remove this global continuum bias, yielding the normalized spectrum $f^\prime_\mathrm{line}$ shown in the middle panel. To eliminate any remaining continuum tilt, we then fit a linear function $f^\prime_\mathrm{linear}$ (red solid line) to $f^\prime_\mathrm{line}$ while masking out the \ion{Mg}{2} absorption feature between $2780$ and $2820 \, \textup{\AA}$. Dividing $f^\prime_\mathrm{line}$ by $f^\prime_\mathrm{linear}$ produces the final corrected profile $\fMg$ in the bottom panel, where the continuum is flat and consistent with unity outside the absorption region. Across all three panels, the color-filled histograms highlight the integrated REW at each stage; for this $r_p$ bin, the two-step correction increases the measured REW by $13\%$ relative to the uncorrected profile $\flin$.

\section{Saturation Correction for REW Measurements}
\label{app:sat_cor}

Deriving the projected cross-correlation function $\wpgc (r_p)$ requires converting the observed total REW into a mean \ion{Mg}{2} column density. As shown by~\citet{Zhu2014}, \ion{Mg}{2} absorption lines begin to saturate at $\WMg \gtrsim 0.15 \, \textup{\AA}$, while weaker absorbers can still exhibit doublet ratios smaller than $2$. Following a similar approach, we calibrate an empirical relation for the doublet ratio $\fdrt$ by combining individual absorbers from the DESI DR2 \ion{Mg}{2} catalog with the weak-absorber sample from~\citet{Narayanan2007}.

\renewcommand\thefigure{B}
\begin{figure}
    \includegraphics[width=\columnwidth]{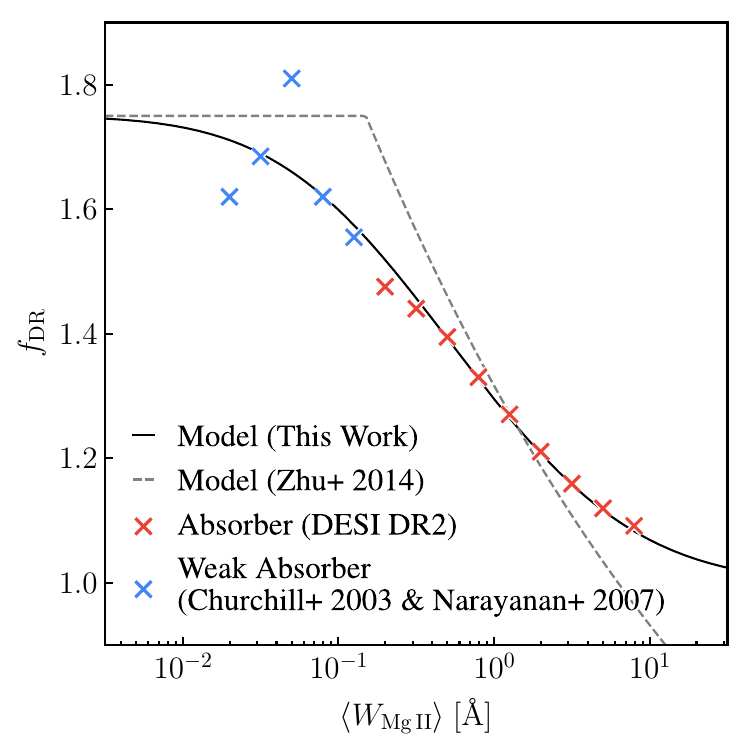}
    \caption{\textit{Doublet ratio as a function of total REW.} Solid red and blue crosses denote the median doublet ratios $\fdrt$ measured from the DESI DR2 \ion{Mg}{2} catalog and the weak-absorber sample of~\citet{Narayanan2007}, respectively. The black solid curve shows our best-fitting doublet ratio model from Equation~\eqref{eqn:dr}, while the dashed gray curve represents the relation adopted by~\citet{Zhu2014}.}
    \label{fig:sat_cor}
\end{figure}

Figure~\ref{fig:sat_cor} presents the binned median doublet ratio $\fdrt$ as a function of total $\langle \WMg \rangle$ measured directly from the DESI DR2 (solid red crosses) and weak-absorber (solid blue crosses) catalogs. Compared to the relation adopted by~\citet{Zhu2014} (dashed gray curve), which assumes a flat optically thin plateau at $\langle \WMg \rangle < 0.15 \, \textup{\AA}$, the combined sample exhibits a more gradual decline in $\fdrt$ with increasing REW and lacks a sharp transition out of the optically thin regime. We model this smooth dependence across the full REW range using the two-parameter function defined in Equation~\eqref{eqn:dr} (black solid curve).

\section{\texorpdfstring{\ion{Mg}{2}}{Mg II} Doublet Discrepancy in the Stacked Measurement of \texorpdfstring{$\xsgc$}{xi}}
\label{app:doublet}

In our main analysis, we derive $\xsgc$ exclusively from the redward half of the stacked $\lambda2803$ absorption profile, as the conversion in Equation~\eqref{eqn:xi^s} assumes optically thin absorption along an unblended wing of the doublet. To test the consistency between the two doublet transitions, we repeat this analysis using the blueward half of the stacked $\lambda2796$ profile.

\renewcommand\thefigure{C}
\begin{figure}
    \includegraphics[width=\columnwidth]{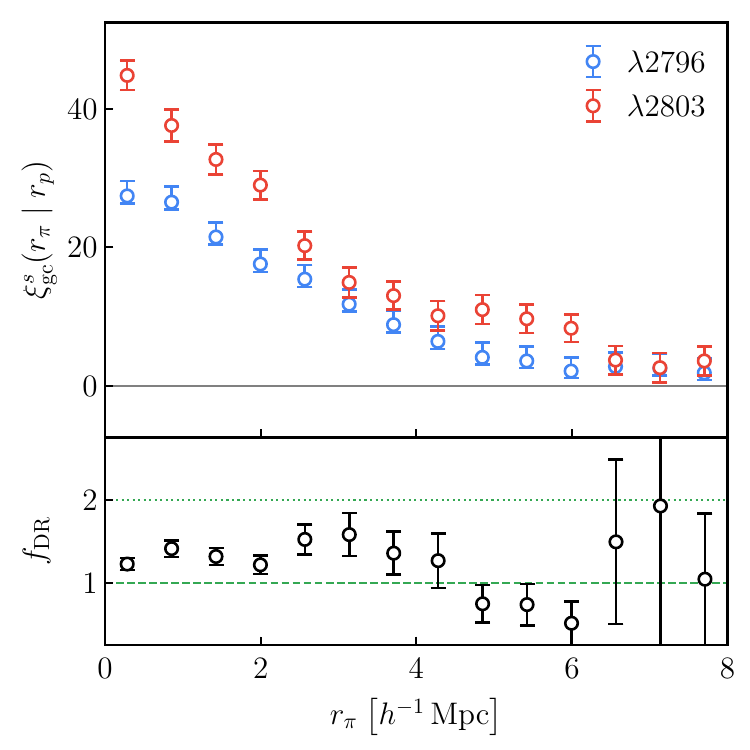}
    \caption{\textit{Stacked $\xsgc$ measurements from the two \ion{Mg}{2} doublet components and their derived doublet ratio.} \textbf{Top:} Redshift-space cross-correlation function $\xsgc$ recovered from the redward $\lambda2803$ (open red circles) and blueward $\lambda2796$ (open blue circles) stacked absorption profiles under the optically thin assumption. \textbf{Bottom:} Derived doublet ratio $\fdrt$ as a function of line-of-sight separation $r_\pi$ (open black circles). Green dotted and dashed horizontal lines indicate the physical limits for optically thin ($\fdrt = 2.0$) and opaque ($\fdrt = 1.0$) absorption, respectively.}
    \label{fig:doublet}
\end{figure}

Figure~\ref{fig:doublet} compares the $\xsgc$ profiles recovered from the $\lambda2803$ and $\lambda2796$ transitions under the optically thin assumption (top panel), along with the derived doublet ratio $\fdrt$ as a function of $r_\pi$ (bottom panel). In the top panel, the $\lambda2796$-based $\xsgc$ profile (open blue circles with errorbars) falls systematically below the $\lambda2803$-based measurement (open red circles with errorbars) across all $r_\pi$ bins, consistent with stronger saturation in the $\lambda2796$ line. In the bottom panel, most of the $\fdrt$ measurements (open black circles with errorbars) lie between the theoretical limits for optically thin ($\fdrt = 2.0$; green dotted line) and opaque ($\fdrt = 1.0$; green dashed line) absorption, but three points drop below unity at $r_\pi = 4.5$--$6.0 \, h^{-1}\mathrm{Mpc}$.

Although two of these three points remain consistent with the physical lower limit of unity within $1\sigma$, this behavior could still raise concern about the validity of the $\lambda2803$-based ``knuckle'' feature. However, several physical mechanisms and independent observations indicate that anomalous \ion{Mg}{2} doublet ratios can arise in dusty or outflowing gas. Using radiative transfer models,~\citet{Prochaska2011} demonstrated that resonant line emission can ``fill in'' absorption troughs, while dust associated with cool gas produces differential reddening that suppresses the redshifted resonant emission. The resulting excess of blueshifted resonant emission would preferentially infill and weaken the blueward stacked $\lambda2796$ absorption relative to the redward $\lambda2803$ wing. Similarly,~\citet{SJChang2025} found stronger $\lambda2803$ than $\lambda2796$ absorption ($\fdrt < 1$) in down-the-barrel spectral stacks of star-forming galaxies, attributing this inverted doublet ratio to intrinsic \ion{Mg}{2} emission entangled with the absorption profiles. More recently, an independent stacking analysis from the MusE GAs FLOw and Wind (MEGAFLOW) survey~\citep{Guo2026} reported ``knuckle''-like features in the redward stacked $\lambda2803$ profile that are likewise absent from the blueward $\lambda2796$ profile.

We also cannot rule out continuum-fitting systematics or statistical fluctuations. For instance,~\citet{PerezRafols2015} showed that certain variable-smoothing methods tend to underestimate the quasar continuum and thus weaken the measured absorption. While our composite spectra show no obvious artifacts such as a broad emission wing, it remains unclear whether a subtle asymmetric continuum bias could suppress the $\lambda2796$ wing more strongly than the $\lambda2803$ wing. By contrast, noise-induced unphysical doublet ratios, which are common in discrete absorber catalogs, should be strongly suppressed when stacking millions of spectra. We test the empirical stability of the $\lambda2803$ absorption signal against observational systematics in the following section.

\section{Robustness Tests of the ``Knuckle''}
\label{app:robust}

As shown in Appendix~\ref{app:doublet}, the ``knuckle'' feature in the 1-halo \ion{Mg}{2} RSD signal appears clearly in the redward $\lambda2803$ wing, whereas the $\lambda2796$ wing and the discrete absorber measurement follow more standard FoG profiles. To test whether the $\lambda2803$ feature is driven by observational systematics or sample variance, we examine its stability across independent datasets and internal subsamples.

\renewcommand\thefigure{D}
\begin{figure*}[ht!]
    \includegraphics[width=\linewidth]{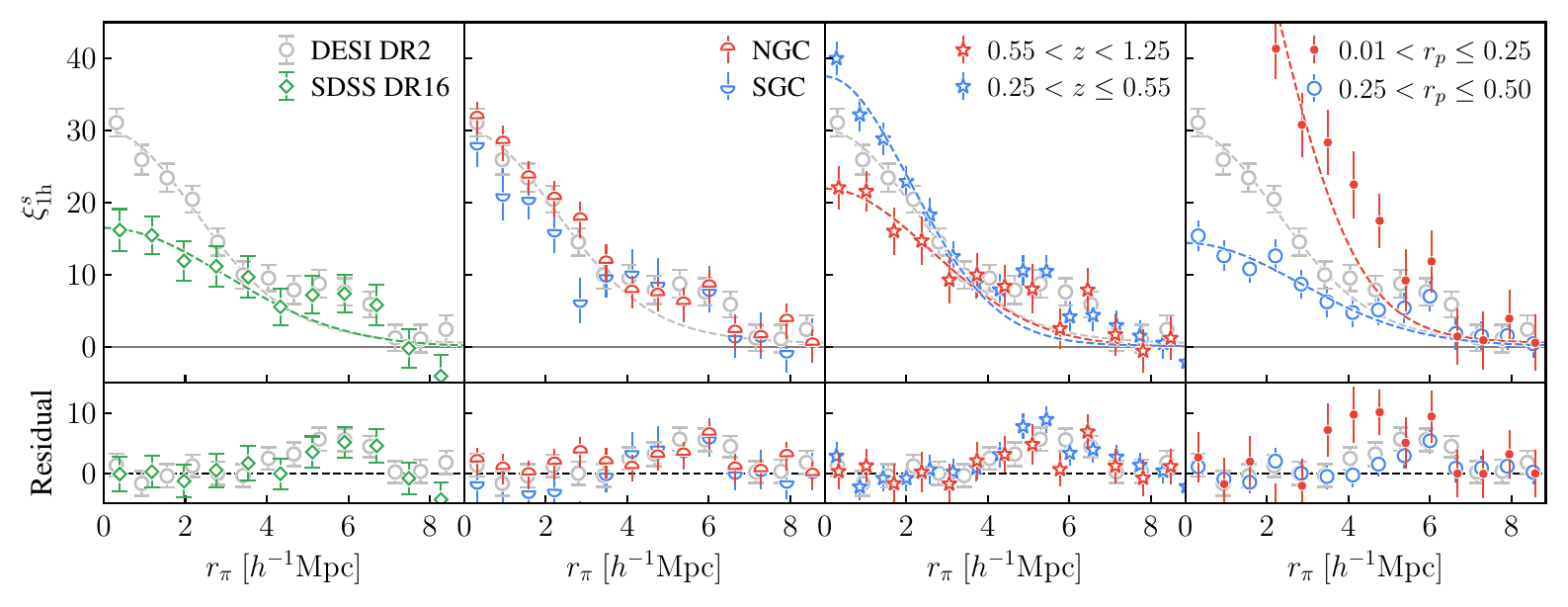}
    \caption{\textit{Robustness tests of the ``knuckle'' feature in the 1-halo \ion{Mg}{2} RSD signal.} \textbf{First column:} The 1-halo RSD signals $\xsOh$ measured from DESI DR2 (gray open circles) and SDSS DR16 (green diamonds). \textbf{Second column:} The DESI DR2 sample split into North Galactic Cap (NGC; blue half circles) and South Galactic Cap (SGC; red half circles) subsamples. \textbf{Third column:} The DESI DR2 sample split into low ($0.25 < z \leq 0.55$; blue stars) and high ($0.55 < z < 1.25$; red stars) redshift subsamples. \textbf{Fourth column:} The DESI DR2 sample split into inner ($0.01 < r_p \leq 0.25 \, h^{-1} \mathrm{Mpc}$; blue circles) and outer ($0.25 < r_p \leq 0.5 \, h^{-1} \mathrm{Mpc}$; red dots) radial subsamples. In all columns, the top panels display the measured $\xsOh$ alongside the best-fitting BST inflow components (dashed curves); for the footprint split (second column), the dashed gray curve represents the reference inflow model derived from the full sample. The bottom panels show the ``knuckle'' residuals after subtracting these inflow components from $\xsOh$. Errorbars indicate $1\sigma$ uncertainties derived from $200$ bootstraps.}
    \label{fig:robust_app}
\end{figure*}

In the first column of Figure~\ref{fig:robust_app}, we compare the $\xsOh$ measurements from DESI DR2 (gray open circles) and SDSS DR16 (green diamonds), along with their respective BST-inferred inflow components (dashed gray and green curves). After subtracting these inflow components in the bottom panel, both the DESI and SDSS residuals exhibit a consistent excess between $4$ and $6 \, h^{-1} \mathrm{Mpc}$, indicating that the ``knuckle'' is unlikely to be an artifact of DESI instrumentation or data reduction.

In the remaining columns of Figure~\ref{fig:robust_app}, we test the internal consistency of the DESI measurement by splitting the sample by survey footprint (NGC vs.\ SGC; second column), redshift ($0.25 < z \leq 0.55$ vs.\ $0.55 < z < 1.25$; third column), and projected separation ($0.01 < r_p \leq 0.25$ vs.\ $0.25 < r_p \leq 0.5 \, h^{-1} \mathrm{Mpc}$; fourth column). Across all six subsamples, the residual excess persists at $> 2\sigma$ significance, albeit with modest variations in shape: the amplitude is somewhat stronger at higher redshifts, and the peak shifts toward lower velocities in the inner radial bin. Together, the agreement between DESI and SDSS and the stability across footprint, redshift, and radial splits indicate that the ``knuckle'' is a robust kinematic feature of the stacked LRG CGM profile rather than an artifact of survey geometry or sample selection.

\section{Recovery of the CGM Velocity Distribution in TNG50}
\label{app:tng_closure}

In \S~\ref{subsec:fog_fit}, we apply the TNG50-motivated BST model to the DESI stacked RSD measurement and show that it reproduces both the central FoG and the secondary ``knuckle'' feature, whereas a single isothermal component fails to capture the high-velocity excess. Here, we perform a closure test on the TNG50 simulation to verify that this deconvolution procedure accurately recovers the underlying 2D velocity distribution of simulated CGM clouds from their 1-halo RSD signal.

\renewcommand\thefigure{E}
\begin{figure*}[ht!]
    \includegraphics[width=\linewidth]{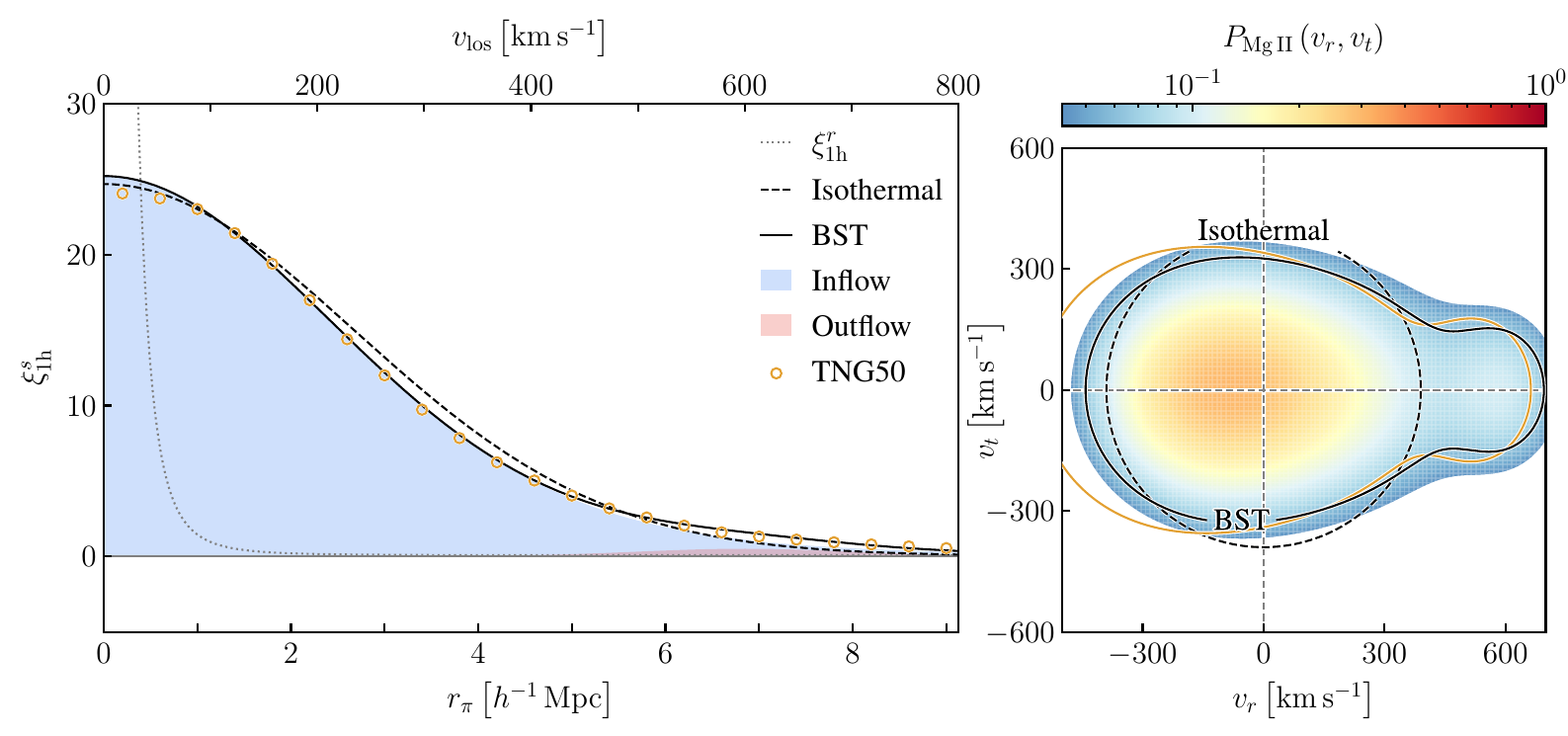}
    \caption{\textit{Recovery of the 2D velocity distribution of cool CGM gas in TNG50.} \textbf{Left:} Orange open circles denote the 1-halo redshift-space LRG-\ion{Mg}{2} cross-correlation function, $\xsOh(r_\pi)$, and the gray dotted curve shows the real-space cross-correlation function, $\xrOh(r)$, both measured directly from TNG50. The solid black curve is the prediction by our best-fitting BST model, decomposed into \emph{Inflow} (blue filled region) and \emph{Outflow} (red filled region) components, while the black dashed curve shows the prediction from a simple isothermal model. \textbf{Right:} Filled color contours and the solid black contour line indicate the 2D VDF of cool CGM gas recovered by the best-fitting BST model, compared against the true 2D VDF measured directly from TNG50 (orange contour line) and the isothermal model prediction (black dashed contour line). All contour lines enclose $75\%$ of the respective populations.}
    \label{fig:fog_tng}
\end{figure*}

We apply the fitting procedure of \S~\ref{subsec:fog_fit} directly to the TNG50 1-halo correlation functions and present the results in Figure~\ref{fig:fog_tng}. In the left panel, the best-fitting BST model (solid black curve), decomposed into its \emph{Inflow} (blue shaded region) and \emph{Outflow} (red shaded region) components, reproduces the simulated $\xsOh(r_\pi)$ profile (orange open circles) over the full $r_\pi$ range. In the right panel, the 2D VDF inferred from this RSD fit (filled contours and solid black contour line) closely matches the true bimodal velocity distribution measured directly from the TNG50 gas cells (orange contour line).

Unlike the DESI stacked RSD profile, the outflowing component in TNG50 manifests in $\xsOh(r_\pi)$ as a broad high-velocity tail rather than a detached ``knuckle''. However, the finite numerical resolution of TNG50 is known to suppress the survival and condensation of cool gas within galactic winds~\citep{Nelson2020}. The absence of a sharp ``knuckle'' in the simulated RSD profile therefore does not conflict with our ``tired wind'' interpretation; rather, both the dynamically tracked outflow population and the extended high-velocity tail in TNG50 support the presence of a secondary outflowing component in LRG halos.

% For this sample we use BibTeX plus aasjournalv7.bst to generate the bibliography. The sample7.bib file was populated from ADS. To get the citations to show in the compiled file do the following:

% pdflatex sample7.tex
% bibtex sample7
% pdflatex sample7.tex
% pdflatex sample7.tex

\bibliographystyle{aasjournalv7}
\bibliography{mgrsd}

\end{document}